\documentclass[pra,twocolumn,
                                superscriptaddress,
                                showpacs,
                                aps
                                ]{revtex4-1}
\usepackage{graphicx}
\usepackage{amsmath}
\usepackage{amsfonts}
\usepackage{amssymb}
\usepackage{hyperref}
\usepackage{color}

\newcommand{\beq}{\begin{equation}}
\newcommand{\eeq}{\end{equation}}

\usepackage{placeins} 
\usepackage{bm}         

\begin{document}

\title{Semiclassical theory of synchronization-assisted cooling}

\author{Simon B. J\"ager} 
\affiliation{Theoretische Physik, Universit\"at des Saarlandes, D-66123 Saarbr\"ucken, Germany} 

\author{Minghui Xu} 
\affiliation{JILA, National Institute of Standards and Technology and Department of Physics,
	University of Colorado, Boulder, Colorado 80309-0440, USA} 
	\affiliation{Center for Theory of Quantum Matter, University of Colorado, Boulder, Colorado 80309, USA} 
\affiliation{Department of Physics and Astronomy, Shanghai Jiao Tong University, Shanghai 200240, China}
\author{Stefan Sch\"utz}
\altaffiliation{present address: icFRC, IPCMS (UMR 7504) and ISIS (UMR 7006), University of Strasbourg and CNRS, 67000 Strasbourg, France.}
\affiliation{Theoretische Physik, Universit\"at des Saarlandes, D-66123 Saarbr\"ucken, Germany}

\author{Murray J. Holland} 
\affiliation{JILA, National Institute of Standards and Technology and Department of Physics,
	University of Colorado, Boulder, Colorado 80309-0440, USA} 
\affiliation{Center for Theory of Quantum Matter, University of Colorado, Boulder, Colorado 80309, USA} 

\author{Giovanna Morigi} 
\affiliation{Theoretische Physik, Universit\"at des Saarlandes, D-66123 Saarbr\"ucken, Germany} 

\date{\today}

\begin{abstract}
We analyse the dynamics leading to radiative cooling of an atomic ensemble confined inside
an optical cavity when the atomic dipolar transitions are incoherently pumped and can synchronize. Our study is performed in
the semiclassical regime and assumes that cavity decay is the largest rate in the system dynamics. 
We identify three regimes characterising the cooling. At first hot atoms are individually cooled
by the cavity friction forces. After this stage, the atoms' center-of-mass motion is further cooled by the coupling to the internal degrees of freedom while the
dipoles synchronize.  In the latest stage dipole-dipole correlations are stationary and the center-of-mass motion is 
determined by the interplay between friction and dispersive forces due to the coupling with the collective dipole. We analyse this asymptotic regime by means of a mean-field model and show that the width of the momentum distribution can be of the order of the photon recoil. Furthermore, the internal excitations oscillate spatially with the cavity standing wave forming an antiferromagnetic-like order. 
\end{abstract}

\maketitle
\section{Introduction}

Radiative cooling is based on tailoring the scattering cross section of photons from atoms, molecules, and optomechanical structures. It achieves a net and irreversible transfer of mechanical energy into the modes of the electromagnetic field by means of a coherent process followed by dissipation, which in atomic and molecular media is usually spontaneous emission \cite{Wineland:1979,Chu:1998}. By these means ultralow temperatures have been realised, paving the way to unprecedented levels of quantum control of the dynamics from the microscopic \cite{Wineland:1999,Metcalf:2003} up to mesoscopic realm \cite{Aspelmeyer,Rubinztein,Eschner:2003}. 

Despite this remarkable progress, radiative cooling of optically-dense atomic or molecular ensembles to quantum degeneracy remains a challenge. Here, cooperative effects of light scattering usually hinder the laser cooling dynamics, because of the enhanced probability of reabsorbing the spontaneously-emitted photons \cite{Walker:1990,Marksteiner:1996,Castin:1998}. Among possible strategies \cite{Cirac:1996} and implementations \cite{Grimm}, one promising scheme uses elastic scattering into the mode of a high-finesse resonator for avoiding spontaneous emission, while the irreversible mechanism leading to dissipation is provided by cavity decay \cite{Horak:1997,Vuletic:2000,Ritsch:RMP,Vuletic:preprint}. In this regime the width of the asymptotic momentum distribution is typically limited by the resonator linewidth \cite{Ritsch:RMP}. In single-mode standing-wave cavities, moreover, the dispersive mechanical forces of the cavity induce a stationary density modulation, which appear when the intensity of the transverse laser driving the atoms exceeds a threshold value \cite{Domokos:2002,Asboth:2005,Schuetz:2014,Schuetz:2015}. 

\begin{figure}[h!]
	\flushleft(a)\vspace{-4ex}\\
	\center\includegraphics[width=0.6\linewidth]{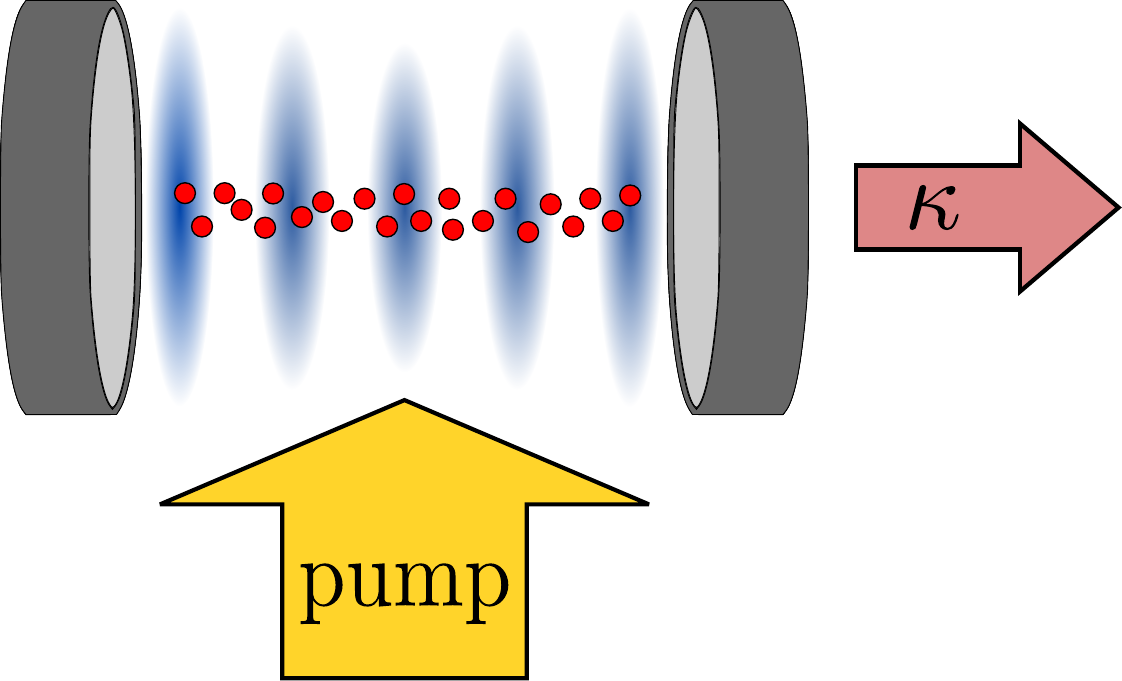}
	\flushleft(b)\vspace{-4ex}\\
	\center \includegraphics[width=0.5\linewidth]{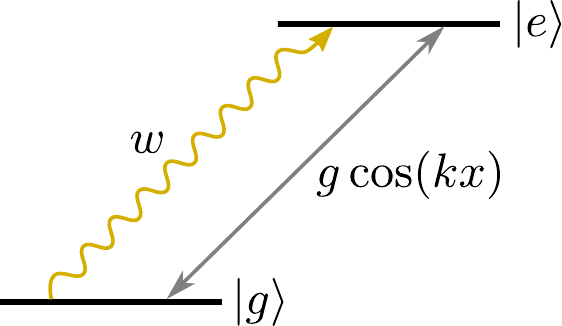}
	\caption{(a) Atoms are transversally driven by an incoherent pump at rate $w$ and strongly couple with the mode of a standing-wave cavity.  (b) The relevant internal states are the two metastable states $|g\rangle$ and $|e\rangle$, which can be the ground and excited states of the intercombination line of an alkali-earth metal atom or two sublevels of a hyperfine multiplet. The dipolar transitions strongly couple to the cavity mode with position-dependent strength $g\cos(kx)$, with $g$ the vacuum Rabi frequency and $k$ the cavity wave number. The cavity decay rate is the largest parameter of the dynamics, {\it i.e.}, $\kappa\gg g, w$.\label{Fig:1}}
\end{figure}

Self-trapping and cooling of atoms in cavities are also expected when the atoms are incoherently pumped \cite{Salzburger:2004,Salzburger:2005,Salzburger:2006}. In setups where the dipoles can synchronize \cite{Bohnet:2012}, a cavity-assisted cooling mechanism was recently identified whose dynamics exhibit giant friction forces \cite{Xu:2016}. Figure \ref{Fig:1} schematically illustrates the setup: the atomic dipolar transitions are transversally driven by an external incoherent pump and strongly couple with the high-finesse mode of a standing-wave resonator, whose decay rate exceeds by orders of magnitude the incoherent pump rate. The numerical analysis performed in Ref. \cite{Xu:2016} showed that the medium could reach ultralow asymptotic temperatures that were orders of magnitude smaller than the cavity linewidth.

\begin{figure}[h!]
	\flushleft(a)\vspace{-4ex}\\
	\hspace{0.4cm}\includegraphics[width=0.95\linewidth]{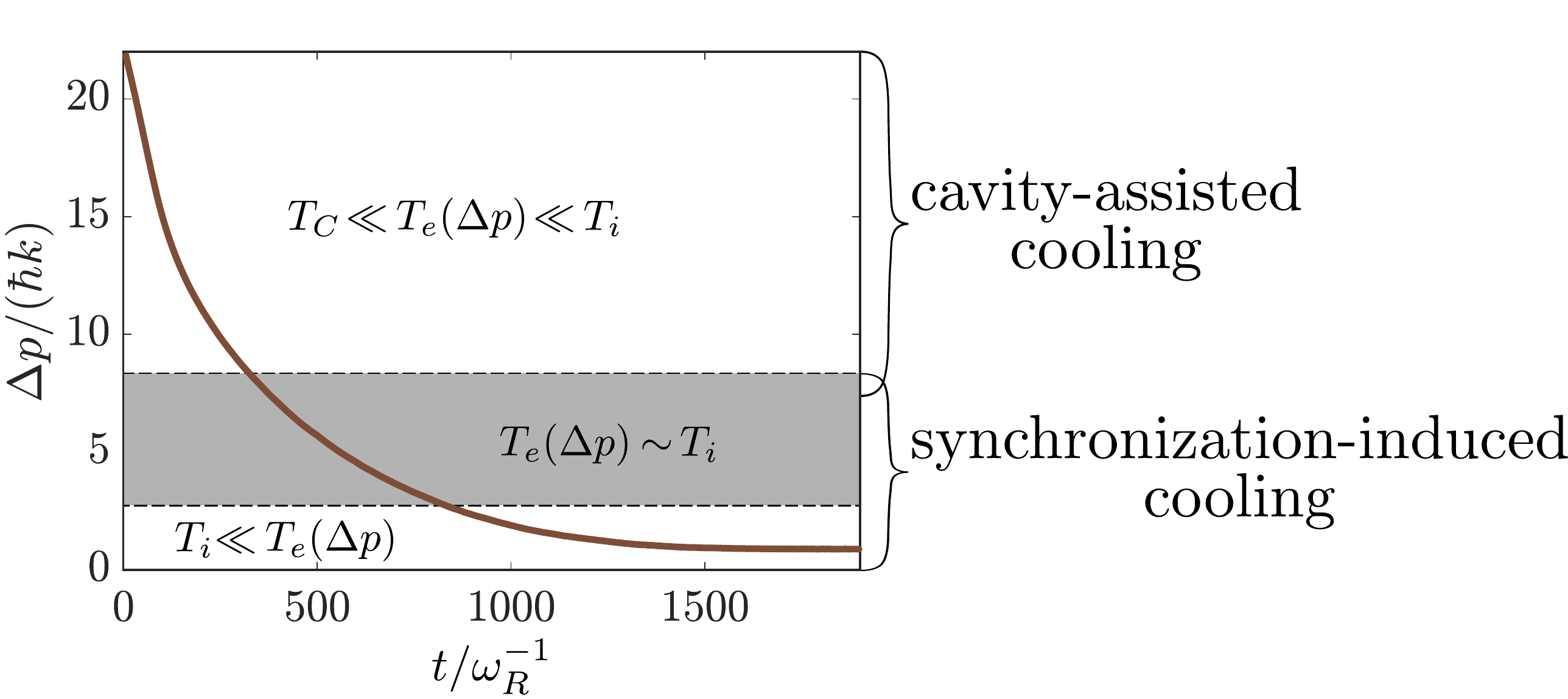}
	\flushleft(b)\vspace{-4ex}\\
	\hspace{0.4cm}\includegraphics[width=0.92\linewidth]{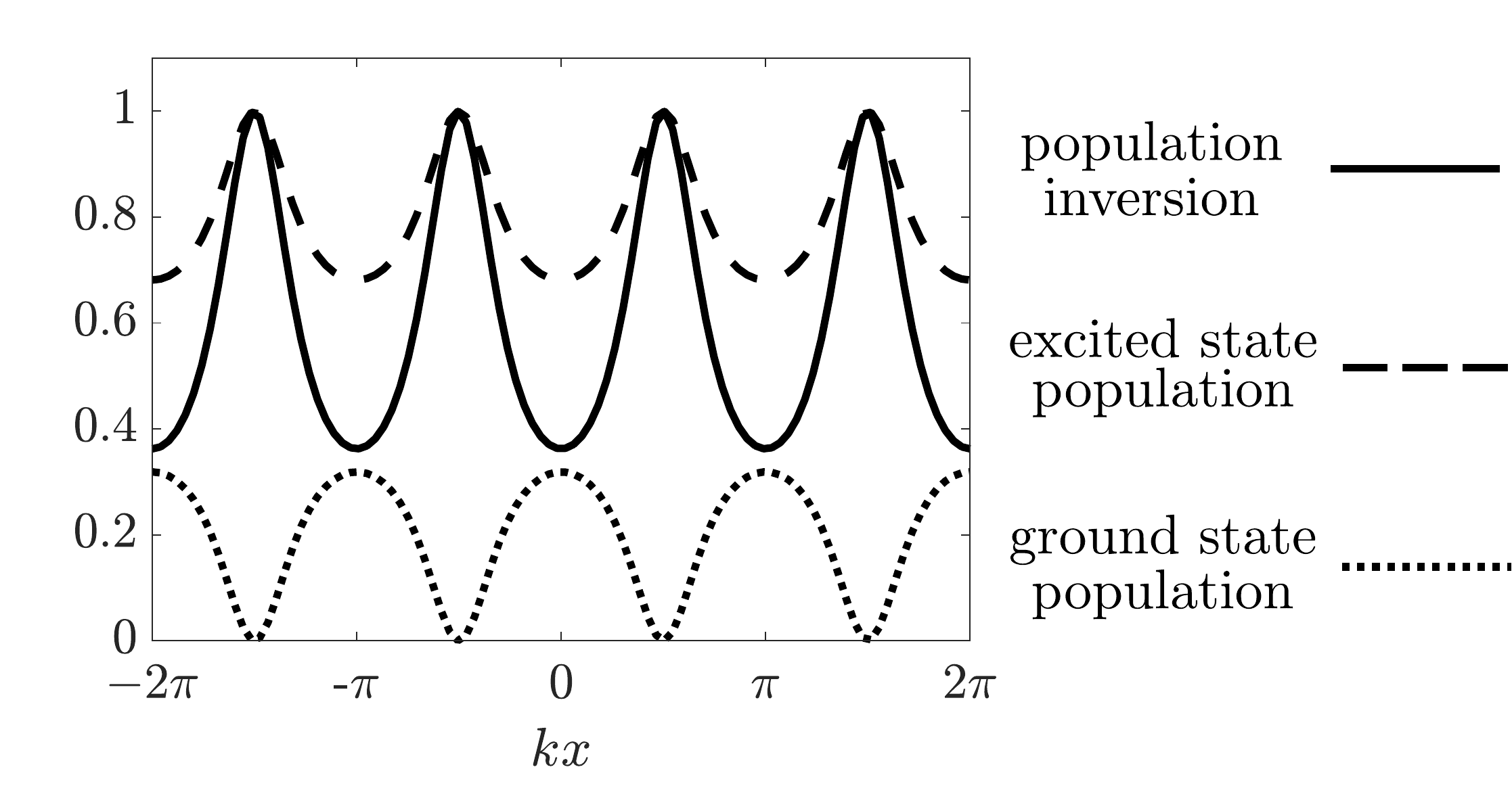}
	\caption{(a) Example of the time evolution of the one-particle momentum width $\Delta p$  and schematic overview of the regimes characterising the dynamics of synchronization-assisted cooling. The width $\Delta p$ determines the characteristic time scale $T_e$ of the atoms' external motion, which is inversely proportional to the mean Doppler shift $k\Delta p/m$. The time scales of reference are determined by cavity decay, $T_C\sim \kappa^{-1}$, and by the spins pump rate, $T_i\sim 1/w$. Initially, $T_i\ll T_e$ and the atoms' center-of-mass motion is cooled by the cavity forces. When the atoms are sufficiently cold that $T_e$ becomes comparable to $T_i$,  the cooling dynamics is determined by the non-adiabatic coupling of the external motion with the spin dynamics. The final stages are characterized by the regime $T_e\gg T_i$ and exhibit temperatures that are orders of magnitude smaller than $\kappa$. (b) The correlations between internal and external degrees of freedom give rise to a position-dependent expectation value of population inversion (solid line) that oscillates with the cavity intensity $\cos^2(kx)$ and is maximum at the nodes. \label{Fig:2}}
\end{figure}

The purpose of this paper is to perform a detailed analysis of the semiclassical dynamics of the synchronization-assisted cooling mechanism of Ref. \cite{Xu:2016}.
Our study extends the work in Ref. \cite{Xu:2016} and builds a consistent theoretical framework from which we can extract analytical predictions on the dynamics. We show that the cooling dynamics is essentially determined by the three stages we illustrate in Fig. \ref{Fig:2}(a): initially hot atoms are cooled by the resonator until the time scale of the external degrees of freedom becomes of the order of the time scale of the internal degrees of freedom. In the second stage the dipoles synchronize and establish correlations with the atoms' spatial distribution. In the final stage dipole-dipole correlations are stationary and the motion is 
cooled down to temperatures that are determined by the pump rate, providing this is chosen within the interval of values allowing synchronization. Even though the steady state exhibits no density modulations, synchronization leads to correlations between the internal and the external degrees of freedom and in the asymptotic limit the atomic excitations oscillate in space with the intensity of the intracavity field, as shown in Fig. \ref{Fig:2}(b). 

The dynamics we discuss complements the studies performed in Refs. \cite{Salzburger:2004,Salzburger:2005,Salzburger:2006}, where the incoherent pump rate was instead the fastest rate of the dynamics. We argue that the resulting regimes are essentially different: for example, in our case at steady state the atoms are not spatially localized and the mean-field character of dipole-dipole correlations is dominant. 


This article is organized as follows. In Sec. \ref{Sec:1} we start from the Heisenberg-Langevin equations of atoms and cavity and derive a semiclassical model for the external degrees of freedom. In Sec. \ref{Sec:2} we determine a mean-field model and test the validity of its predictions. We then use the mean-field model to analyse the steady state and estimate the final temperature. The conclusions are drawn in Sec. \ref{Sec:3}, while the appendices provide details of the calculations of Secs. \ref{Sec:1} and \ref{Sec:2}.

\section{Semiclassical model of Synchronization-induced cooling}
\label{Sec:1}

The system we consider consists of $N$ atoms of mass $m$ that are confined within a high-finesse optical resonator and are constrained to move only along the cavity axis, which we denote by the $x$-axis. The setup is illustrated in Fig. \ref{Fig:1}. The atomic dipolar transition is incoherently driven by a transverse pump (directed orthogonal to the cavity axis) at rate $w$. Each atom is composed of two metastable states, $|g\rangle$ and $|e\rangle$, and strongly couples to a mode of the cavity with position-dependent strength $g\cos(kx)$, with $g$ the vacuum Rabi frequency and $k$ the cavity wave number. We discard the instability of the excited state, so that atomic emission only occurs into the cavity mode. This can be realized when $|g\rangle$ and $|e\rangle$ form the intercombination line of alkali-earth metals \cite{Meiser:2009} or when they are two substates of the hyperfine multiplet coupled by a two-photon transition, of which one dipole transition is coupled with the resonator \cite{Bohnet:2012}. In either case the atomic transition frequency $\omega_a$ is determined by the energy splitting between the two levels, while the mechanical effects of light scale with the recoil frequency $\omega_R=\hbar k^2/(2m)$. To good approximation, this is determined by the wave number $k$ of the cavity mode. 

In this section we start from the Heisenberg-Langevin equations of motion for the cavity, electronic, and center-of-mass degrees of freedom, and derive the equations in the limit in which the atoms' center-of-mass motion can be treated semiclassically. The parameter regime we consider is the one of synchronization: The cavity decay rate $\kappa$ is the fastest rate of the dynamics and the value of the incoherent pump rate $w$ is chosen within the lower and the upper synchronization thresholds \cite{Meiser:2010:1}, as we specify below. The recoil frequency $\omega_R$ is typically the smallest parameter of the dynamics, so that $\omega_R<w\ll\kappa$, which is consistent with the validity of the semiclassical treatment we apply in this work. 

\subsection{Heisenberg-Langevin equations}

We denote by $\hat a$ and $\hat a^\dagger$ the annihilation and creation operators of a cavity photon at frequency $\omega_c$ and wave number $k$. The atoms are assumed to be distinguishable and are labeled by $j$ ($j=1,\ldots,N$). Their canonically-conjugated position and momentum are denoted by $\hat{x}_j$ and  $\hat{p}_j$ and the lowering, raising, and population-inversion operators by $\hat{\sigma}_j=|g\rangle_j\langle e|$, $\hat{\sigma}_j^{\dag}=|e\rangle_j\langle g|$, and $\hat{\sigma}^{z}_j=|e\rangle_j\langle e|-|g\rangle_j\langle g|$, respectively. 

The Hamiltonian governing the coherent dynamics in the frame rotating at the frequency $\omega_a$ reads
\begin{align}
\hat{H}=\hbar\Delta\hat{a}^\dagger\hat{a}+
\sum_{j=1}^N\frac{\hat{p}_j^2}{2m}
+\hbar\frac{g}{2}\sum_{j=1}^{N}(\hat{a}^\dagger
\hat{\sigma}_j\cos(k\hat{x}_j)+\mathrm{H.c.})\,,
\end{align}
with $\Delta=\omega_c-\omega_a$ the detuning between cavity and atomic transition frequency.

The Heisenberg-Langevin equations for the relevant operators include the cavity damping at rate $\kappa$, the incoherent pump at rate $w$, and the corresponding Gaussian input noise operators $\hat{a}_{\text{in}}$ and $\hat{\sigma}_{\text{in},j}$, respectively, and read
\begin{align}
\frac{d}{dt}\hat{x}_{j}=&\frac{\hat{p}_j}{m},\label{quantumx}\\
\frac{d}{dt}\hat{p}_{j}=&\hbar k \frac{g}{2}(\hat{a}^{\dag}\hat{\sigma}_{j}\sin(k\hat{x}_j)+\mathrm{H.c.}),\label{quantumforce}\\
\frac{d}{dt}\hat{\sigma}_{j}=&-\frac{w}{2}\hat{\sigma}_j+i\frac{g}{2}\cos(k\hat{x}_j)\hat{\sigma}_j^{z}\hat{a}-\sqrt{w}\hat{\sigma}_{j}^{z}\hat{\sigma}_{\text{in},j}^{\dag}, \label{quantumspin}\\
\frac{d}{dt}\hat{\sigma}_{j}^{z}=&w(1-\hat{\sigma}_j^{z})+\left(ig\hat{a}^{\dag}\hat{\sigma}_{j}\cos(k\hat{x}_j)+\mathrm{H.c.}\right)\nonumber\\
&+2\sqrt{w}\left(\hat{\sigma}_{\text{in},j}\hat{\sigma}_j+\hat{\sigma}_j^{\dag}\hat{\sigma}_{\text{in},j}^{\dag}\right), \label{quantumspinz}\\
\frac{d}{dt}\hat{a}=&\left(-i\Delta-\frac{\kappa}{2}\right)\hat{a}-\sum_{j=1}^Ni\frac{g}{2}\hat{\sigma}_j\cos(k\hat{x}_j)+\sqrt{\kappa}\hat{a}_{\text{in}}.\label{quantuma}
\end{align}
Here, $\langle \hat{\sigma}_{\text{in},j}(t)\rangle=0=\langle \hat{a}_{\text{in}}(t)\rangle$, $\langle \hat{\sigma}_{\text{in},j}^{\dag}(t)\hat{\sigma}_{\text{in},j'}(t')\rangle=0=\langle \hat{a}_{\text{in}}^{\dag}(t)\hat{a}_{\text{in}}(t')\rangle$, $\langle \hat{a}_{\text{in}}(t)\hat{a}_{\text{in}}^{\dag}(t')\rangle=\delta(t-t')$, and $\langle \hat{\sigma}_{\text{in},j}(t)\hat{\sigma}_{\text{in},j'}^{\dag}(t')\rangle=\delta_{jj'}\delta(t-t')$. The expectation values $\langle\cdot\rangle$ are taken over the tensor product between the initial density matrix of system and external Markovian environment with vanishing mean number of photons \cite{Gardiner:1985}. 

\subsection{Coarse-grained dynamics}

We derive an effective model by assuming that the decay rate of the resonator $\kappa$ is the largest rate of the dynamics. This allows us to identify a coarse-grained time scale $\Delta t$ that is infinitesimal for the internal degrees of freedom but over which the cavity degrees of freedom can be eliminated from the equations of the atomic dynamics. 

The coarse-grained cavity field operator is given by the time-average $\bar{\hat{a}}(t)=\langle \hat{a}(t)\rangle_{\Delta t}$, where $\langle \hat{\zeta}(t)\rangle_{\Delta t}\equiv\int_t^{t+\Delta t}{\rm d}t'\hat{\zeta}(t')/\Delta t$. It
takes the form
\begin{align}
\bar{\hat{a}}(t)\approx\frac{-i\frac{Ng}{2}}{\kappa/2+i\Delta}\left(\left\langle\hat{X}\right\rangle_{\Delta t}-\frac{1}{\kappa/2+i\Delta}\left\langle\frac{d}{dt}\hat{X}\right\rangle_{\Delta t}\right)+\hat{\mathcal{F}}(t)\,,\label{eliminateda:main}
\end{align}
and is here expressed  in terms of the synchronization order parameter $\hat{X}$ of Ref. \cite{Xu:2016}:
\begin{align}
\hat{X}(t)=\frac{1}{N}\sum_{l=1}^N \hat{\sigma}_l\cos(k\hat{x}_l)\label{X}\,.
\end{align}
It is possible to provide a physical interpretation of the various terms on the right-hand side (RHS) of Eq. \eqref{eliminateda:main}. The first term is the adiabatic component, where the cavity field follows instantaneously the atomic state given by $\hat X(t)$. The second term depends on the time derivative of $\hat X$, and thus on memory effects of the internal and external degrees of freedom in lowest order. It is hence a non-adiabatic correction. Finally, the third term on the RHS gives the contribution of the quantum noise, whose explicit form is reported in Appendix \ref{App:A}. 

Before we proceed, we observe that the characteristic time scales of the atomic motion, and thus of $\hat X$, are determined by the incoherent pump rate $w$ for the internal degrees of freedom, $T_i\sim 1/w$, and by the mean kinetic energy $E_{\rm kin}=\langle p^2\rangle/(2m)$ for the external degrees of freedom, as illustrated in Fig. \ref{Fig:2}(a). More specifically, the characteristic time of the external motion scales with $T_e\sim 1/R_{\rm Doppler}$, where $R_{\rm Doppler}\approx 2\sqrt{\omega_R E_{\rm kin}/\hbar}$ is the mean Doppler shift. When the atoms are sufficiently hot, it is necessary to include non-adiabatic corrections when eliminating the cavity field. 
However, since $T_i$ is typically orders of magnitude larger than $T_C$ for the parameters of interest, only the retardation effects of the external degrees of freedom can be relevant over the time scale $\Delta t$, hence in Eq. \eqref{eliminateda:main} we shall use
\begin{align}
\label{X:nonad}
\left\langle\frac{d}{dt}\hat{X}\right\rangle_{\Delta t}\approx \frac{1}{N}\sum_{j=1}^N\left\langle\hat{\sigma}_j\frac{d}{dt}\left(\cos(k\hat{x}_j)\right)\right\rangle_{\Delta t}\,.
\end{align}

On the basis of these considerations, in the coarse-grained time scale the dynamics of the internal degrees of freedom is solely determined by the adiabatic component of the cavity field. The corresponding equations read (from now on the operators are assumed to be in the coarse-grained time scale and we omit to write $\langle\, \cdot\,\rangle_{\Delta t}$):
\begin{align}
\frac{d}{dt}\hat{\sigma}_{j}=&-\frac{w}{2}\hat{\sigma}_j+\frac{N\Gamma_C}{2}(-i\alpha^*)\cos(k\hat{x}_j)\hat{\sigma}_j^{z}\hat{X}\nonumber\\
&-\sqrt{w}\hat{\sigma}_{j}^{z}\hat{\sigma}_{\text{in},j}^{\dag},\label{sigma2}\\
\frac{d}{dt}\hat{\sigma}_{j}^{z}=&w(1-\hat{\sigma}_j^{z})-\left(N\Gamma_{C}(i\alpha)\hat{X}^{\dag}\hat{\sigma}_j\cos(k\hat{x}_j)+\mathrm{H.c.}\right)\nonumber\\
&+2\sqrt{w}\left(\hat{\sigma}_{\text{in},j}\hat{\sigma}_j+\hat{\sigma}_j^{\dag}\hat{\sigma}_{\text{in},j}^{\dag}\right)\label{sigmaz2}\,,
\end{align}
where 
\begin{align}
\label{Gamma:C}
\Gamma_C=&\frac{g^2/4}{\Delta^2+\kappa^2/4}\kappa\,
\end{align} 
is the effective atomic linewidth while 
\begin{align*}
\alpha=\frac{\Delta}{\kappa/2}-i\,
\end{align*} 
is a dimensionless parameter, which is purely imaginary when $\Delta=0$. 

Retardation effects are instead important for the dynamics of the external degrees of freedom in the initial stage of the dynamics. By keeping the non-adiabatic corrections to the cavity field, according to the prescription of Eq. \eqref{X:nonad}, their dynamics read
\begin{align}
\frac{d}{dt}\hat{p}_j &= \hat{F}_j^{(0)}+\hat{F}_{j}^{(1)}+\hat{\mathcal{N}}_j\label{fullforce}\,.
\end{align}
Here, $\hat{F}_j^{(0)}$ and $\hat{F}_j^{(1)}$ are force operators that describe the adiabatic and non-adiabatic contribution of the cavity field, respectively, while $\hat{\mathcal{N}}_j$ describes a position-dependent Gaussian noise:
\begin{align}
\hat{F}_j^{(0)}=&-\hbar k\frac{N\Gamma_C}{2}\alpha\hat{X}^{\dag}\hat{\sigma}_j\sin(k\hat{x}_j)+\mathrm{H.c.}\label{F0}\\
\hat{F}_j^{(1)}=&-\frac{N\Gamma_C}{2}\frac{\omega_R\kappa}{\Delta^2+\kappa^2/4}i\alpha^2\label{frictionpart}\\
\times&\frac{1}{2N}\sum_{l=1}^N\left[\sin(k\hat{x}_l),\hat{p}_l\right]_{+}\hat{\sigma}_l^{\dag}\hat{\sigma}_j\sin(k\hat{x}_j)+\text{H.c.}\,\nonumber\\
\hat{\mathcal{N}}_j=&\hbar k\frac{g}{2}\hat{\mathcal{F}}(t)^{\dag}\hat{\sigma}_j\sin(k\hat{x}_j)+\text{H.c.}\,,
\label{noise}
\end{align}
where $\left[A,B\right]_{+}=AB+BA$.  The non-adiabatic component $\hat{F}_j^{(1)}$ scales with the ratio $R_{\rm Doppler}/\sqrt{\kappa^2/4+\Delta^2}$ with respect to the adiabatic component. It can be discarded in the later stages of the dynamics, that is when the atoms are sufficiently cold so that $R_{\rm Doppler}\sim w$ corresponding to $T_e\sim T_i$. 

It is important to emphasize the motivation for the definition of the synchronization order parameter $\hat{X}$ in Eq. \eqref{X}. This definition generalizes the collective spin $\hat{j}_{-}=\left(\sum_{i=1}^N\hat{\sigma}_i\right)/N$, which has a non vanishing expectation value in the synchronized phase \cite{Meiser:2010:1}. Operators $\hat{X}$ and $\hat{j}_{-}$ in fact coincide when the atoms are localized at the positions where $\langle\cos(k\hat{x}_j)\rangle=1$ for all atoms. In the generalized form, the synchronization order parameter $\hat{X}$ depends explicitly on the correlations between the internal degrees of freedom and the atomic positions within the cavity optical lattice. We will see that this property can lead to cooling when the dipoles synchronize.

\subsection{Semiclassical dynamics of the external degrees of freedom}

We now assume that the width $\Delta p$ of the single atom momentum distribution is $\Delta p\gg\hbar k$ at all stages of the dynamics, where $\hbar k$ is the linear momentum carried by a cavity photon. In this limit the recoil frequency $\omega_R$ is assumed to be the smallest frequency scale and a semiclassical description of the atomic center-of-mass motion is justified \cite{Stenholm:1986}. By means of this description the equation of motion for the atomic momentum reads
\begin{align}
\frac{d}{dt}p_j=F+\xi_j^p\,,\label{scfull}
\end{align}
where $F=F_j^{(0),\text{sc}}+F_j^{(1),\text{sc}}$ and
\begin{align}
F_j^{(0),\text{sc}}=&-\hbar k\sin(kx_j)\frac{N\Gamma_C}{2}\alpha\hat{X}^{\dag}\hat{\sigma}_j+\mathrm{H.c.}\,\label{classicalforce},\\
F_j^{(1),\text{sc}}=&-\frac{N\Gamma_C}{2}\frac{\omega_R\kappa}{\Delta^2+\kappa^2/4}i\alpha^2\sin(kx_j)\nonumber\\
\times&\frac{1}{N}\sum_{l=1}^N\sin(kx_l)p_l\hat{\sigma}_l^{\dag}\hat{\sigma}_j+\text{H.c.}\,.\label{classicalFriction}
\end{align}
The stochastic variable $\xi_j^p$ describes the properties of the Gaussian noise, $\langle\xi_j^p(t)\xi_l^p(t')\rangle=D^{jl}\delta(t-t')$ with
\begin{align}
D^{jl}=\Gamma_C\hbar^2k^2\sin(kx_j)\sin(kx_l)\mathrm{Re}[\langle
\hat{\sigma}_l^{\dag}\hat{\sigma}_j\rangle]\,.
\end{align}
According to this semiclassical model, the dynamics is determined by these equations together with the equations $\dot x_j=p_j/m$  and the quantum mechanical equations for the internal degrees of freedom, Eqs. \eqref{sigma2} and \eqref{sigmaz2}. The latter, in particular, now depend on the semiclassical variables $x_j$.

Figures \ref{tempcomparison} and \ref{X2comparison} display $\langle p^2\rangle$ and the corresponding correlations $\langle X^\dagger X\rangle$ as a function of time, assuming that there are initially no correlations between the dipoles and that at $t=0$ the atoms' motion is in a thermal state at a given temperature $T$ (the subplots from top to bottom correspond to decreasing values of $T$). The solid curves have been numerically evaluated by integrating Eqs. \eqref{sigma2}, \eqref{sigmaz2},  after performing a second order cumulant expansion for the spins as shown in Appendix \ref{App:B}, together with the stochastic differential equation \eqref{scfull} \cite{SDE}. For the parameter choice we considered the semiclassical dynamics predict the exponential decrease of the kinetic energy towards an asymptotic value which is of the order of the recoil energy. Comparison with the time-evolution of the correlations $\langle X^\dagger X\rangle$  show that these reach the asymptotic value at a rate comparable with the initial cooling rate. These correlations can be measured by detecting the  intracavity photon number, since $\langle \hat{a}^{\dag}\hat{a}\rangle\propto N^2\langle \hat{X}^{\dag}\hat{X}\rangle$, and signify the build-up of spin-spin correlations and of correlations between the spins and  their external positions within the cavity lattice. We denote by synchronization-induced cooling the cooling dynamics that is intrinsically connected with the buildup of these correlations and thus of the intracavity field.

\subsection{Unravelling the semiclassical dynamics}

\begin{figure}[h!]
	\flushleft(a)\vspace{-4ex}\\
	\center\includegraphics[width=0.7\linewidth]{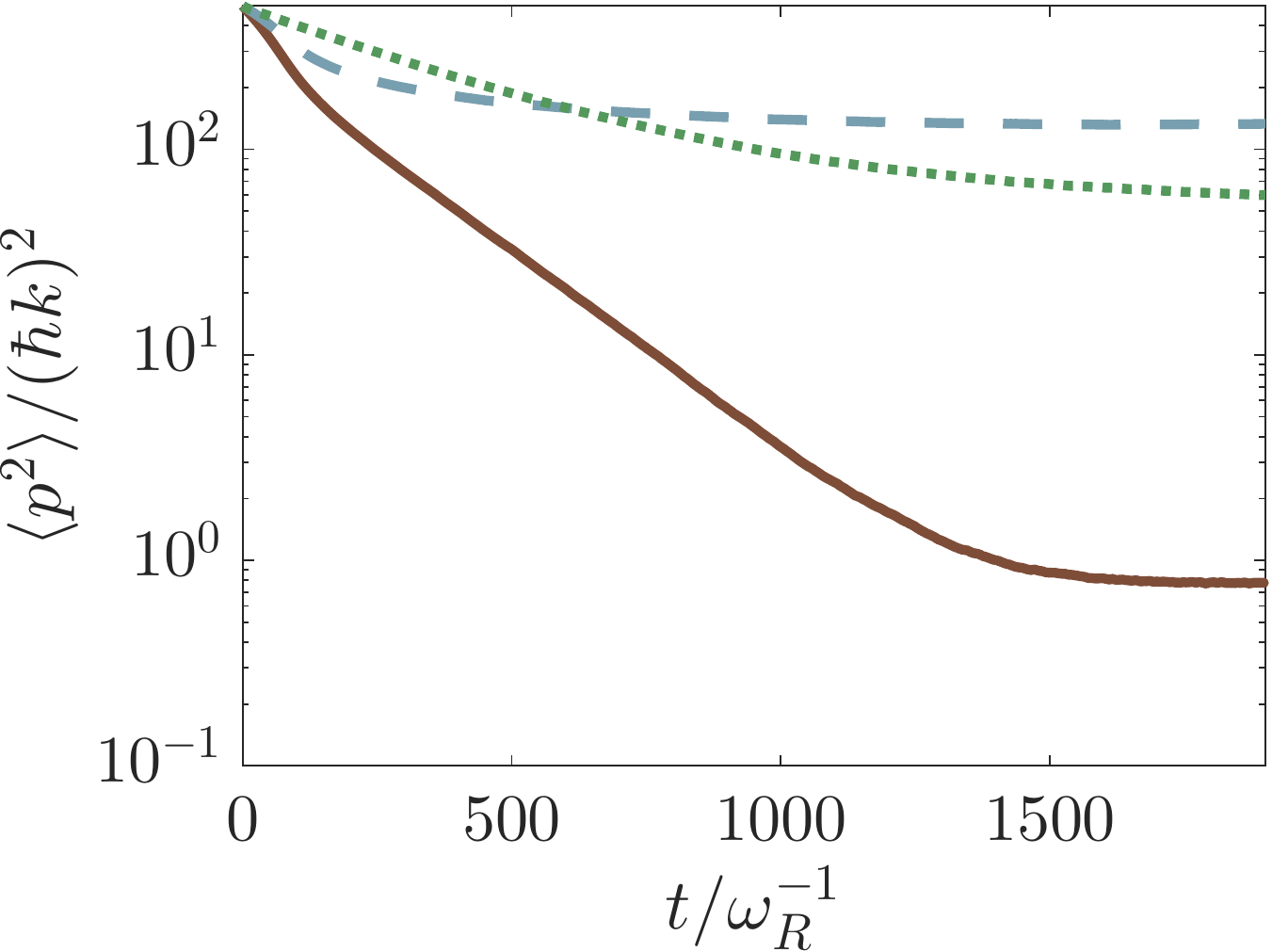}
	\flushleft(b)\vspace{-4ex}\\
	\center\includegraphics[width=0.7\linewidth]{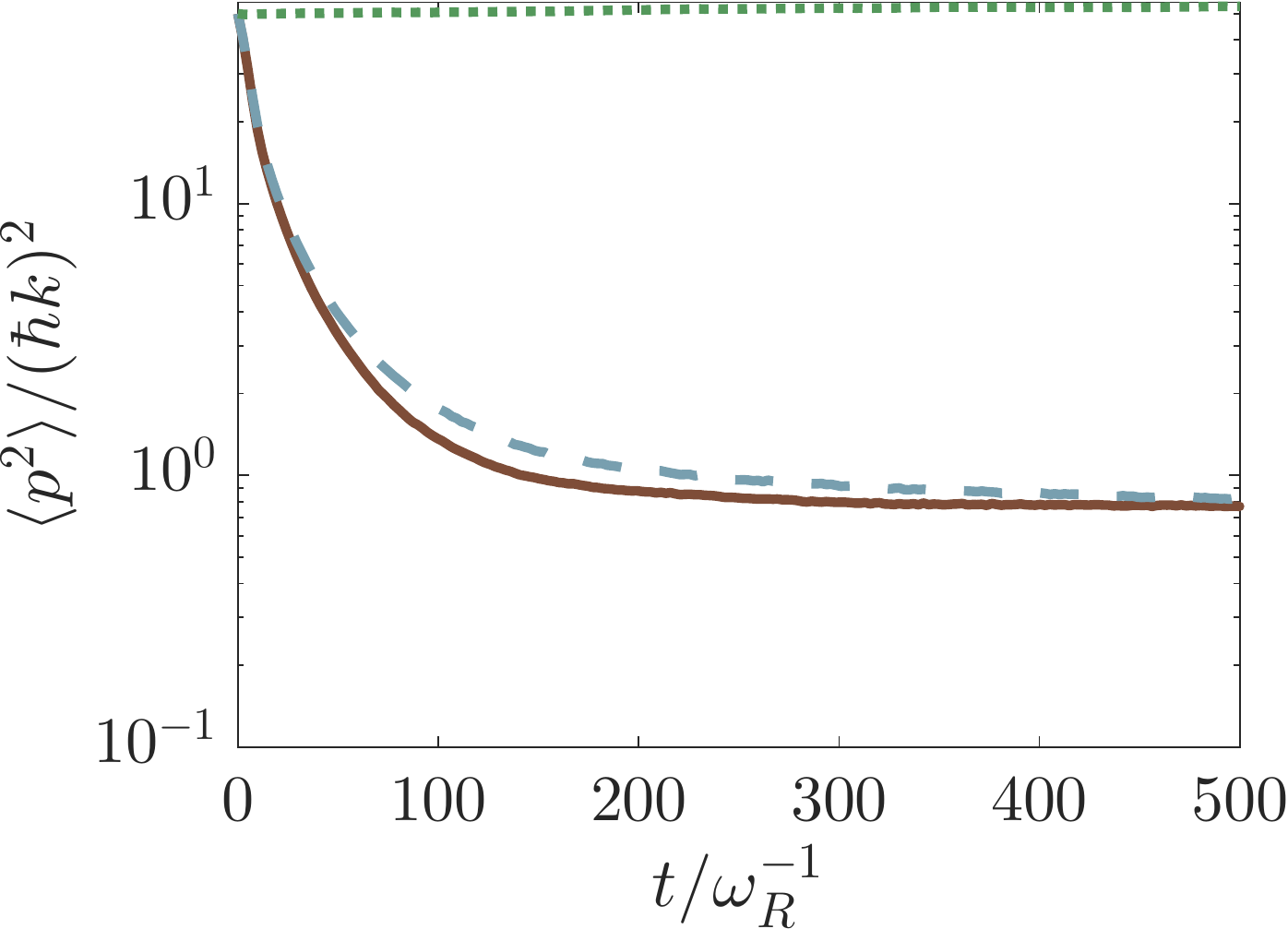}
	\flushleft(c)\vspace{-4ex}\\
	\center\includegraphics[width=0.7\linewidth]{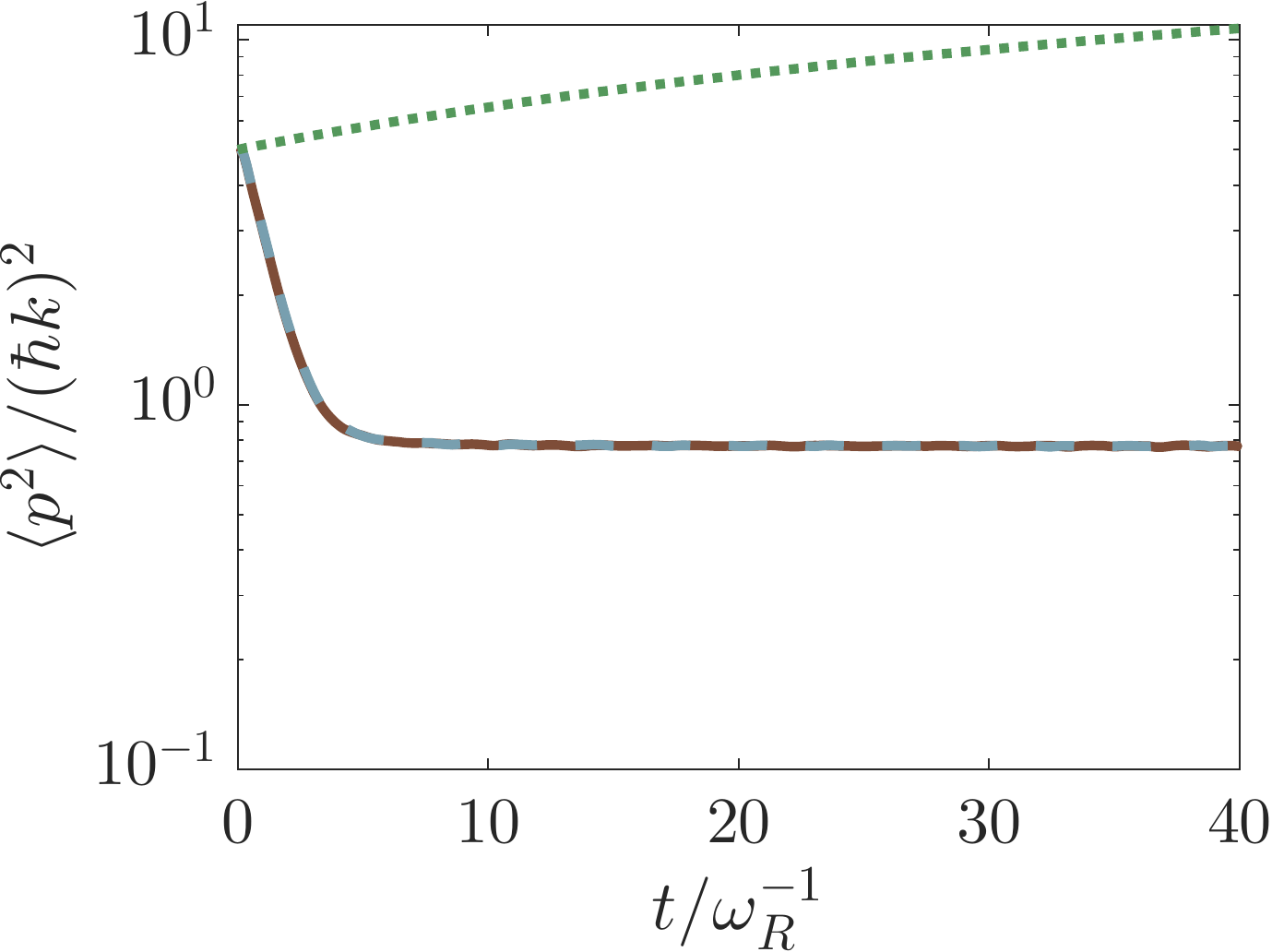}
	\caption{(color online) Dynamics of the width $\langle p^2\rangle$ of the single-atom momentum distribution (in units of $(\hbar k)^2$) as a function of time (in units of $1/\omega_R$). The solid curves are determined by integrating Eqs. \eqref{sigma2}, \eqref{sigmaz2} (using the cumulant expansion, see Appendix \ref{App:B}) and  \eqref{scfull} assuming that initially all atoms are in the excited state and are uniformly spatially distributed, while their momentum distribution is thermal with width (a) $\langle p^2(0)\rangle=500(\hbar k)^2$; (b) $\langle p^2(0)\rangle=50(\hbar k)^2$; (c) $\langle p^2(0)\rangle=5(\hbar k)^2$. The dashed and dotted lines are the corresponding simulations obtained by integrating the equations after setting in Eq. \eqref{scfull} $F=F_j^{(1),\text{sc}}$(dotted line) and $F=F_j^{(0),\text{sc}}$ (dashed line). The parameters are $N=100$, $\kappa=780\omega_R$, $N\Gamma_C=40\omega_R$, $\Delta=\kappa/2$, $w=N\Gamma_C/4$.}\label{tempcomparison}
\end{figure}

\begin{figure}[h!]
	\flushleft(a)\vspace{-4ex}\\
	\center\includegraphics[width=0.7\linewidth]{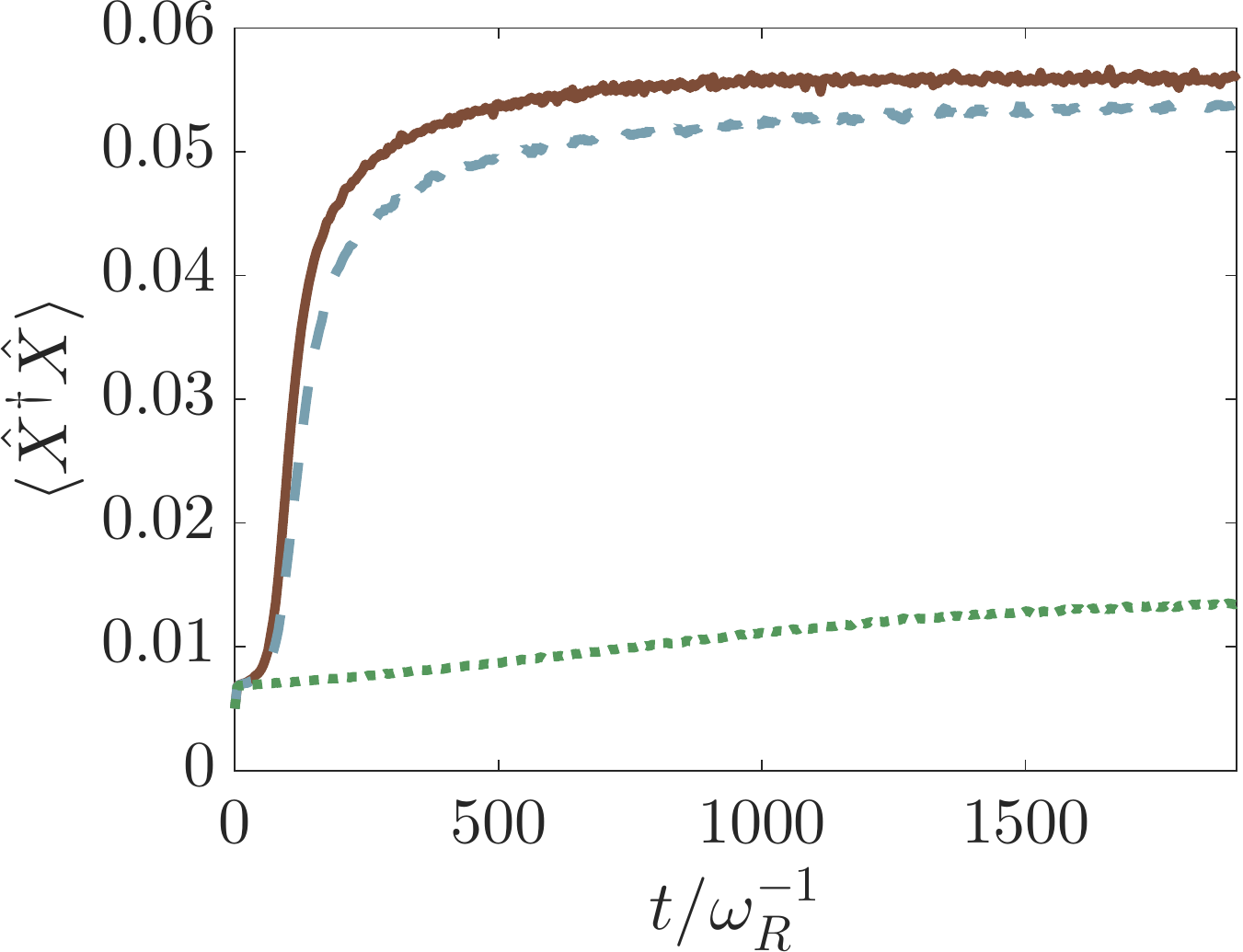}
	\flushleft(b)\vspace{-4ex}\\
	\center\includegraphics[width=0.7\linewidth]{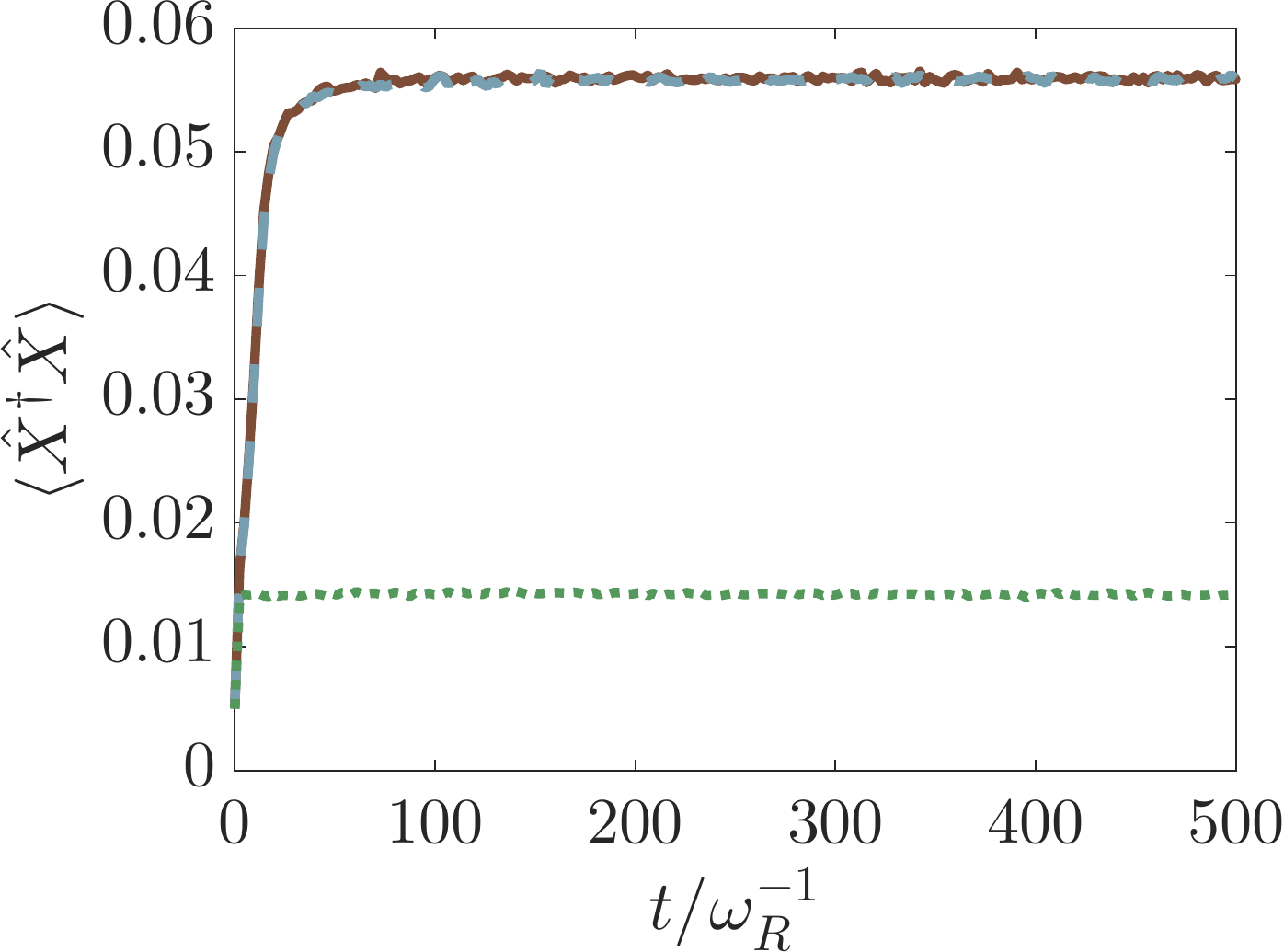}
	\flushleft(c)\vspace{-4ex}\\
	\center\includegraphics[width=0.7\linewidth]{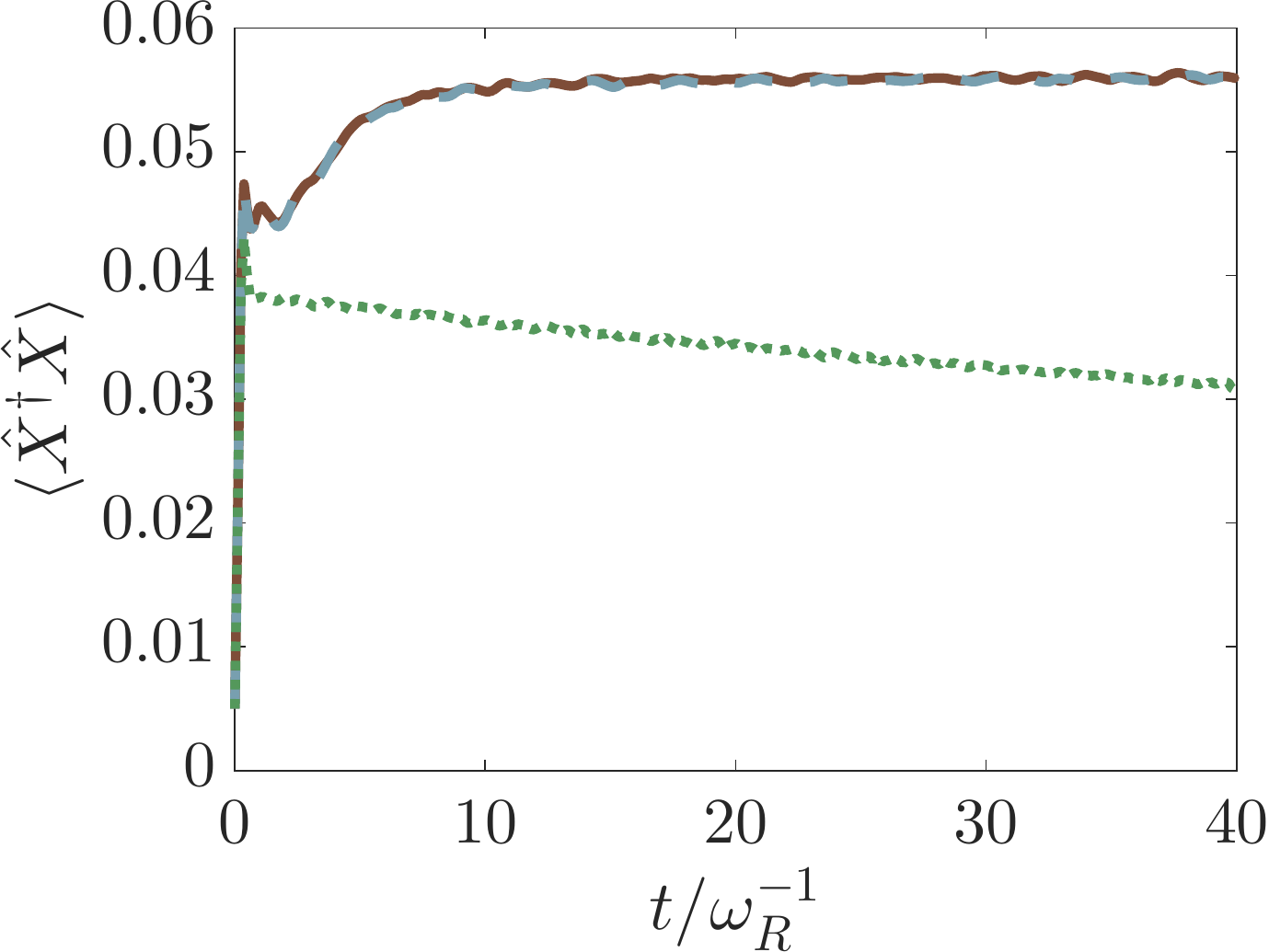}
	\caption{(color online) Correlation $\langle\hat{X}^{\dag}\hat{X}\rangle$ as a function of time (in units of $1/\omega_R$). This quantity signifies the occurrence of synchronization. Subplots (a)-(c) respectively correspond to the dynamics of the subplots (a)-(c) of Fig. \ref{tempcomparison}.}\label{X2comparison}
\end{figure}

In order to gain insight into the mechanisms which lead to the observed behaviour, we compare these curves with the corresponding predictions obtained by either only considering the cavity friction component of the force given in Eq. \eqref{classicalFriction}, thus setting $F=F_j^{(1),\text{sc}}$ in Eq. \eqref{scfull} (dotted line), or by only considering the component of  the force given in Eq.\eqref{classicalforce}, thus setting $F=F_j^{(0),\text{sc}}$ in Eq. \eqref{scfull} (dashed line). The resulting curves in Fig. \ref{tempcomparison}(b)-(c) and of Fig. \ref{X2comparison}(b)-(c) show that $F_j^{(0),\text{sc}}$ is primarily responsible for the build-up of spin-spin correlations and for the cooling dynamics when the atomic initial temperature is sufficiently low (the friction force tends instead to heat the distribution in (c)). 

\begin{figure}[h!]
	\flushleft(a)\vspace{-4ex}\\
	\center\includegraphics[width=0.8\linewidth]{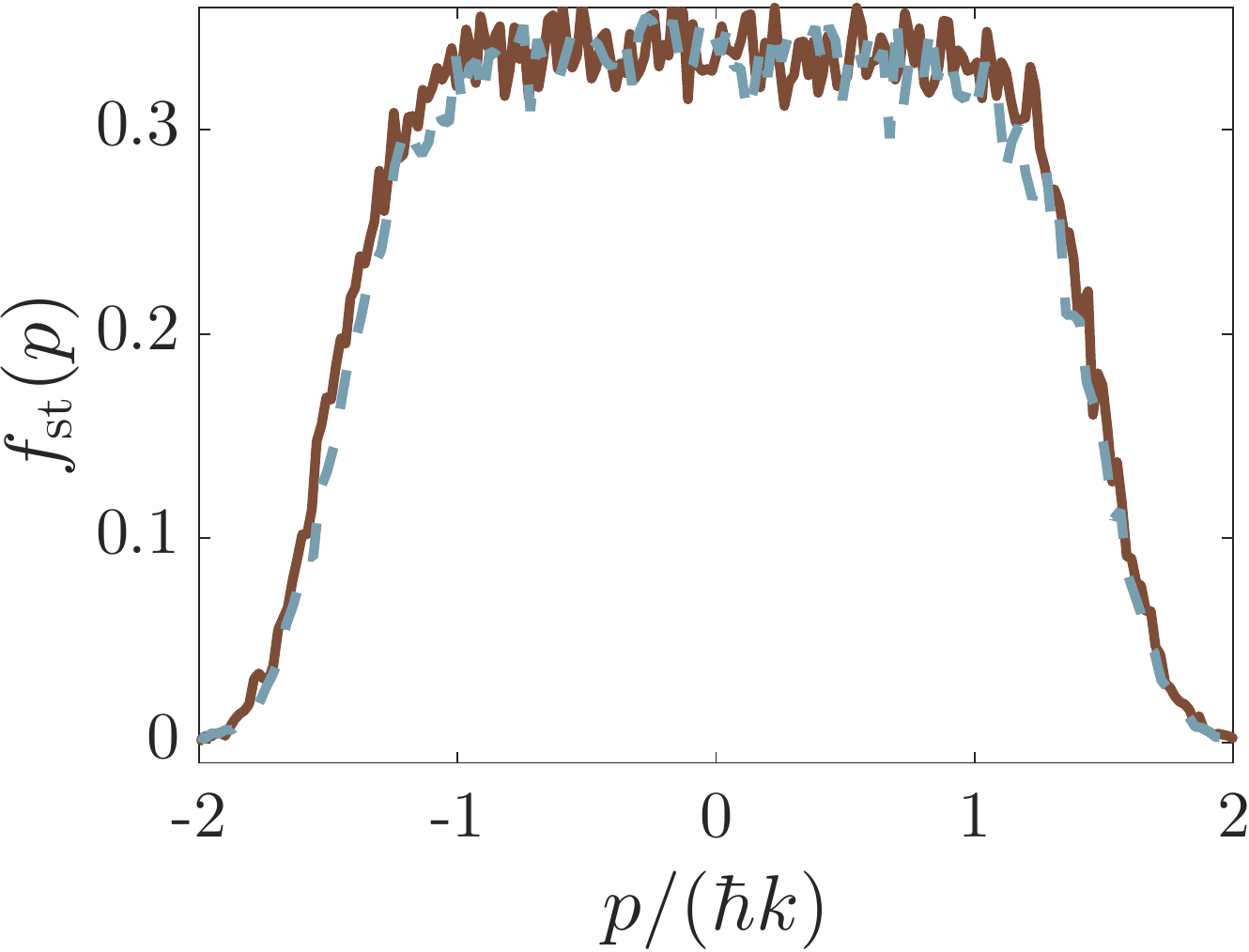}
	\flushleft(b)\vspace{-4ex}\\
	\center \includegraphics[width=0.8\linewidth]{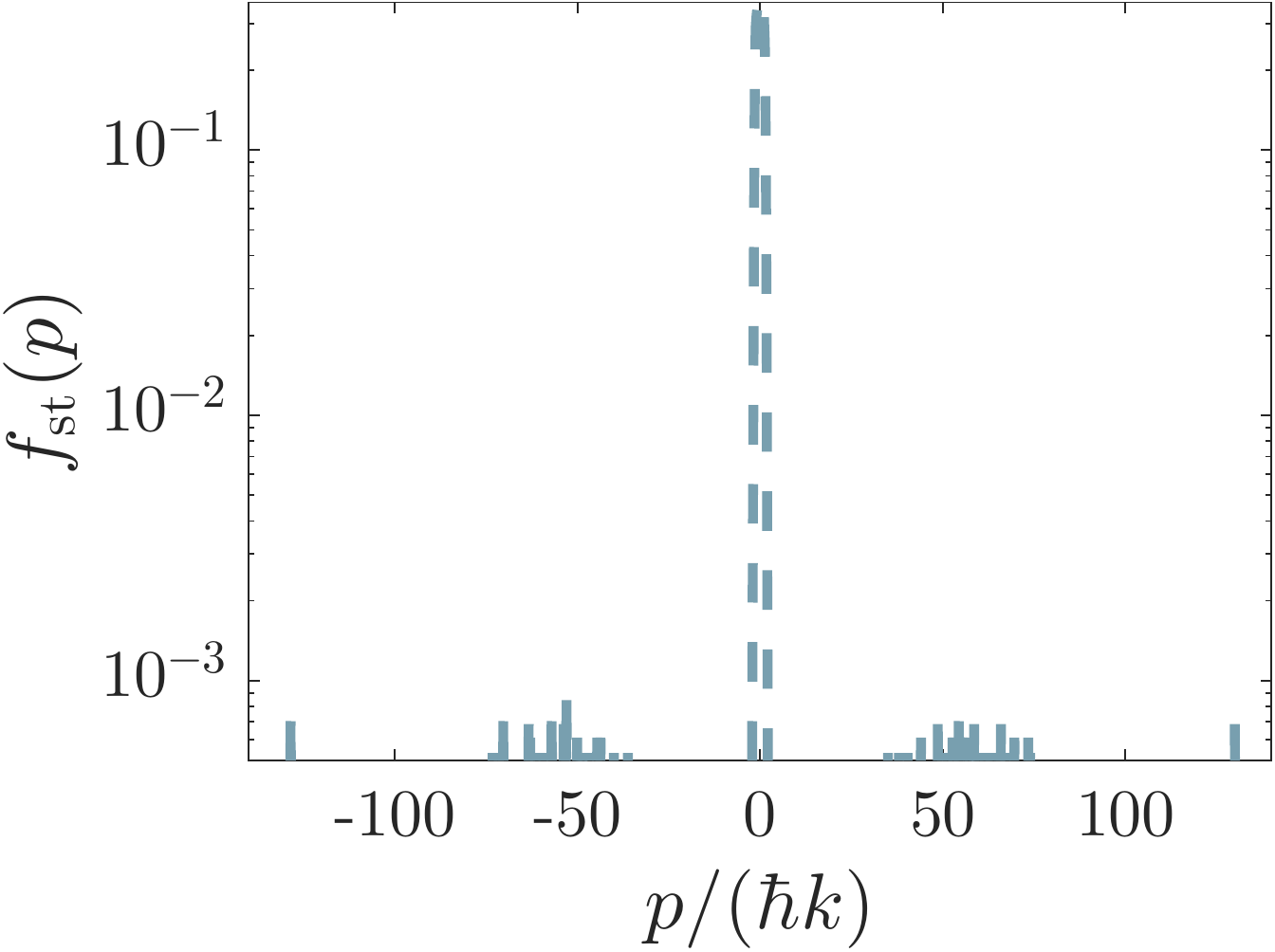}
	\caption{(a) Momentum distribution $f_{\text{st}}(p)$ as a function of $p$ (in units of $\hbar k$) and resulting from the dynamics of Fig. \ref{tempcomparison} (a) at the time $t\simeq 2000\omega_R^{-1}$. The solid and dashed lines illustrate the momentum distributions obtained by considering the full force and only the adiabatic component, respectively, in Eq. \eqref{scfull}. Subplot (b) shows the momentum distribution (in logarithmic scale) over the full initial range of values, demonstrating the existence of long tails. These are responsible for the discrepancy observed in the asymptotic limit of the corresponding curves in Fig.  \ref{tempcomparison} (a). }
	\label{momentumdist}
\end{figure}

A different behaviour is observed in Figs. \ref{tempcomparison}(a) and \ref{X2comparison}(a): Although the friction force contributes to the build up of the cavity field, none of the individual components reproduce the full semiclassical dynamics. In particular, the adiabatic component leads to a larger asymptotic value of the mean kinetic energy, while the cavity friction force cools the motion at a significantly slower rate. Figure \ref{momentumdist} displays the momentum distribution that each of these dynamics predict at $t\approx 2\cdot10^3\omega_R^{-1}$. Remarkably, they qualitatively agree for small momenta, as visible in subplot (a). In (b), however, we observe discrepancies at large momenta. The distribution due to the adiabatic component of the force exhibits atoms at large momenta. These atoms, instead, are cooled by the cavity friction force. We further note that the momentum distribution is approximately flat in the momentum interval $[-\hbar k,\hbar k]$, suggesting that the stationary state is non-thermal.

\begin{figure}[h!]
	\center\includegraphics[width=0.8\linewidth]{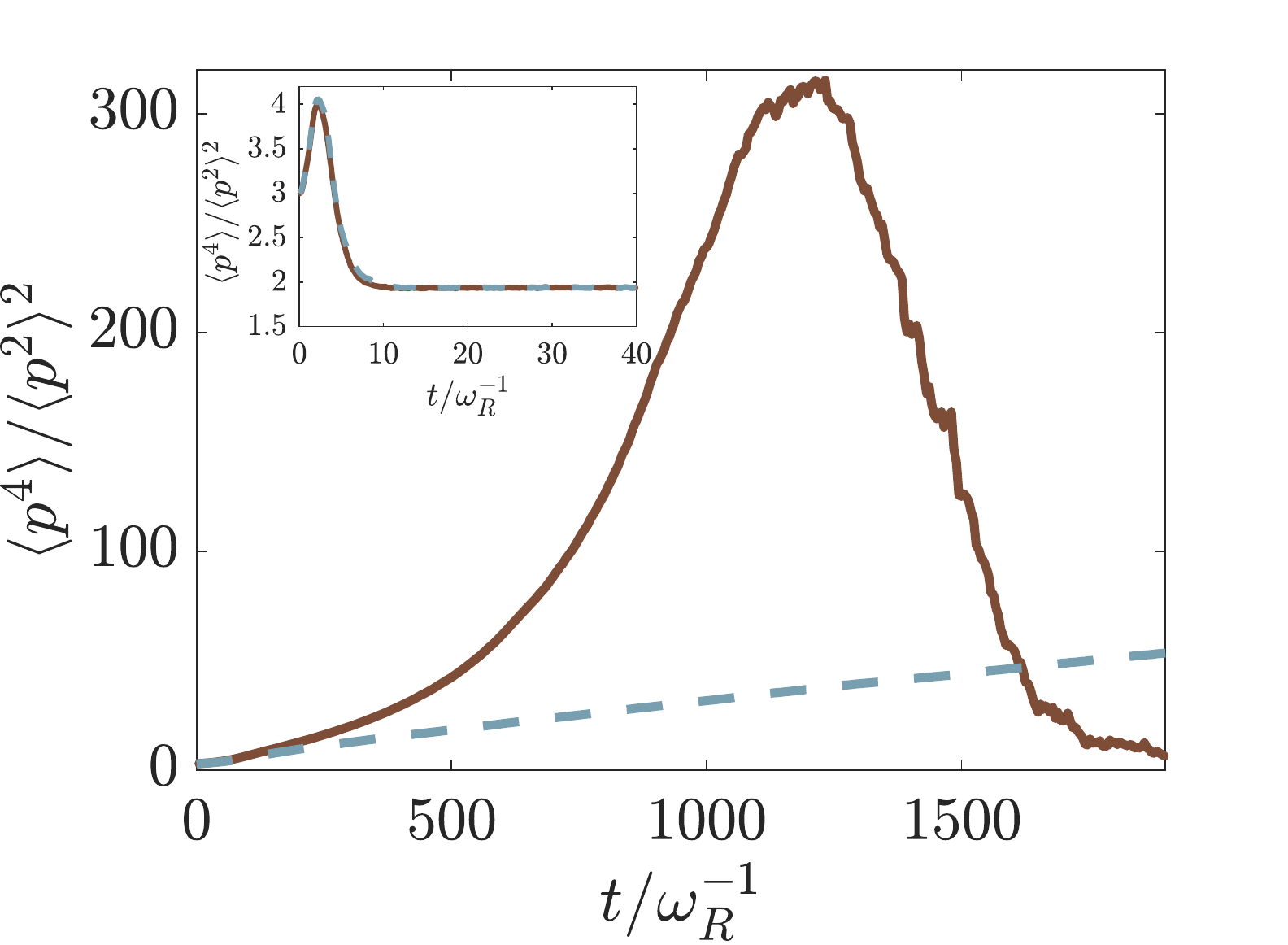}
	\caption{Kurtosis $\mathcal K=\langle p^4\rangle/\langle p^2\rangle^2$ as a function of time for the dynamics of Fig.  \ref{tempcomparison} (a). The inset shows the kurtosis for the parameters as in Fig. \ref{tempcomparison} (c): Both curves relax to approximately the same value $\mathcal K\approx2$.}
	\label{Fig:K}
\end{figure}

We further characterize the dynamics by inspecting the time evolution of the Kurtosis $\mathcal K(t)$, which is  defined as $$\mathcal K(t)=\langle p(t)^4\rangle/\langle p(t)^2\rangle^2\,,$$ with $\langle p(t)^n\rangle$ the $n$-th moment of the single particle distribution at time $t$. The Kurtosis for a Gaussian distribution is 3, so that deviations from this value signal that the distribution is non-thermal. Figure \ref{Fig:K} shows the Kurtosis $\mathcal K(t)$ for the dynamics reported in Fig. \ref{tempcomparison}(a). The distribution is non-thermal at all times, including the asymptotic limit,  where it tends towards the value 2. The large value it reaches during the evolution is attributed to the existence of tails of the momentum distribution at large $p$. These components are cooled by the cavity friction force at a later stage of the dynamics, as visible by comparing these dynamics with the one in which the cavity friction force is set to zero (dashed line). When instead the initial temperature is very low, the Kurtosis is well described by the sole effect of the adiabatic component of the force, as visible in the inset of Fig. \ref{Fig:K}.

This analysis suggests that the friction and the adiabatic forces have very different velocity capture ranges, and in particular the cavity friction forces precool the atoms until a regime in which retardation effects become very small. In this regime, we will show that synchronization-induced cooling efficiently concentrates the atoms in a narrow velocity distribution that can be of the order of the recoil frequency. The time scales associated with these dynamics are illustrated in Fig. \ref{Fig:2} and are at the basis of the theoretical treatment presented in what follows.


\section{Local mean-field model}
\label{Sec:2}

We now analyse the dynamics in the regime where the dipoles have synchronized, corresponding to the stage where the correlations $\langle X^\dagger X\rangle$ have built up. We perform our study by
means of a mean-field approximation, namely, by assuming 
\begin{eqnarray}
&&\langle \hat{\sigma}_j\rangle=s_j\,,\\
&&\langle \hat{\sigma}^{z}_j\rangle=z_j\,,
\end{eqnarray}
where $s_j,z_j$ are scalars. This consists of approximating $\langle \hat{\sigma}_j^{\dag}\hat{\sigma}_i\rangle\approx s_j^*s_i$ for $i\neq j$. Within this treatment, the synchronization order parameter reads
\begin{equation}
X=\langle \hat{X}\rangle=\sum_{j=1}^Ns_j\cos(kx_j)/N\,.
\end{equation}
It is worth emphasizing that we keep the correlations between the internal and the external degrees of freedom, but assume that particle-particle correlations are of mean-field type.

In the mean-field approximation Eqs. \eqref{sigma2} and \eqref{sigmaz2} take the form
\begin{align}
\frac{ds_j}{dt}=&-\frac{w}{2}s_j-\frac{N\Gamma_C}{2}i\alpha^*X\cos(kx_j)z_j,\label{sj}\\
\frac{dz_j}{dt}=&w(1-z_j)+2N\Gamma_C\mathrm{Im}\left\{\alpha X^{*}s_j\right\}\cos(kx_j),\label{zj}
\end{align}
where the noise due the incoherent pump is neglected. The corresponding mean-field equations for the external degrees of freedom read:
\begin{align}
\frac{dx_j}{dt}=&\frac{p_j}{m}, \label{ballistic}\\
\frac{dp_j}{dt}=&-\hbar k\sin(kx_j)N\Gamma_C\mathrm{Re}\left\{\alpha X^*s_j\right\}\,,\label{meanfieldforce}
\end{align}
where the force is the adiabatic component, Eq. \eqref{classicalforce}, and consistent with the mean-field treatment we have discarded cavity shot noise. 

\subsection{Comparison between mean-field and semiclassical model}

We now test the predictions of the mean-field equations by comparing the mean-field dynamics with ones obtained integrating the semiclassical equations, Eqs. \eqref{sigma2}, \eqref{sigmaz2}, and \eqref{scfull}. Since the mean-field treatment should more faithfully reproduce the full dynamics for increasing number of particles, we perform simulations for $N=100$ and $N=1000$ particles. In doing so we rescale the coupling strength $g$ so to keep $Ng^2$ and thus $N\Gamma_C$ constant (compare with Eq. \eqref{Gamma:C}). This implies that the upper synchronization threshold, $w=N\Gamma_C$ \cite{Meiser:2010:1}, is a constant for this thermodynamics limit, while the lower threshold $w=\Gamma_C$ \cite{Meiser:2010:1} in this case scales with $1/N$ and thus vanishes for $N\to\infty$. 

\begin{figure}[h!]
	\flushleft(a)\vspace{-4ex}\\
	\center \includegraphics[width=0.8\linewidth]{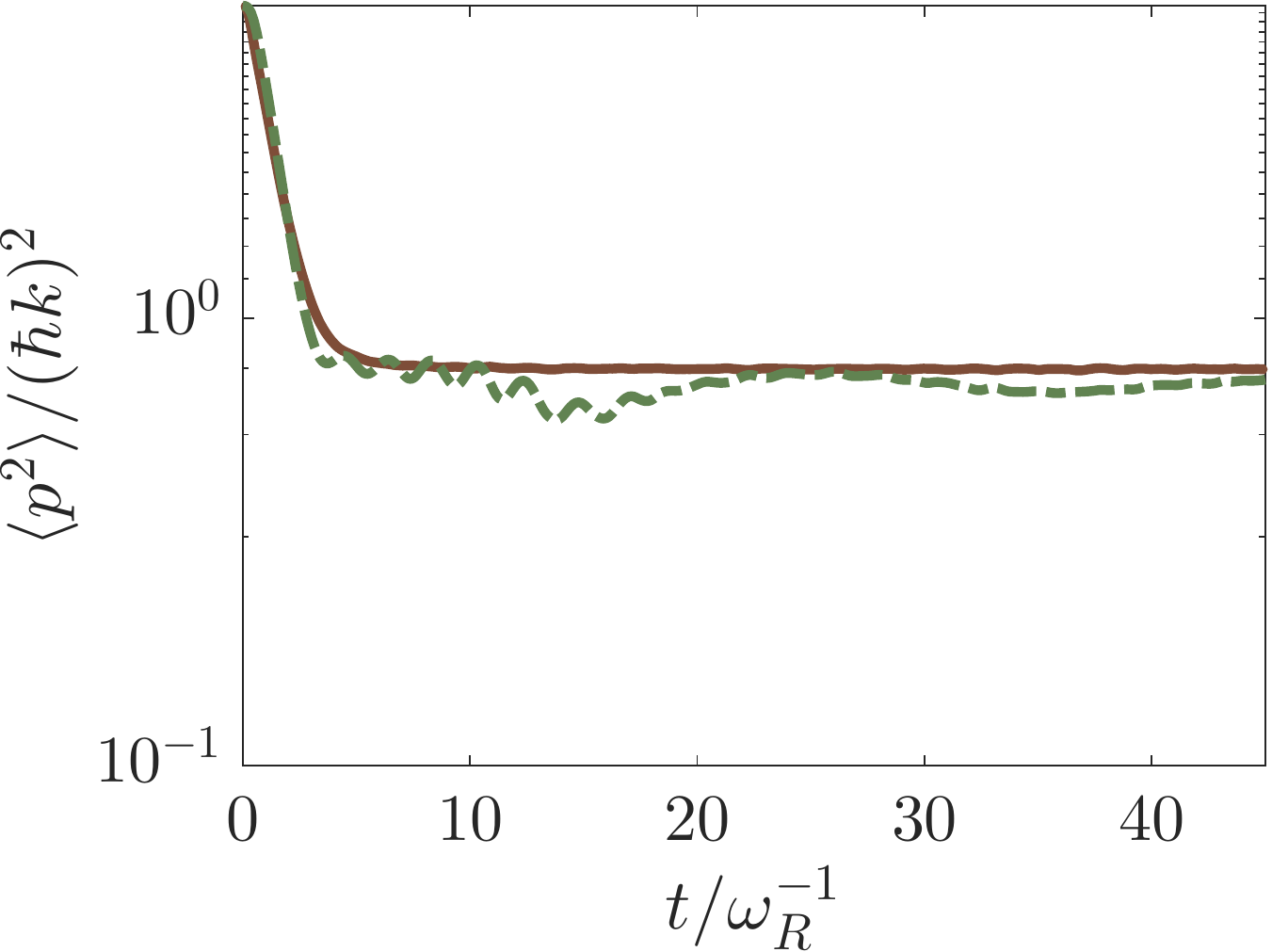}
	\flushleft(b)\vspace{-4ex}\\
	\center \includegraphics[width=0.8\linewidth]{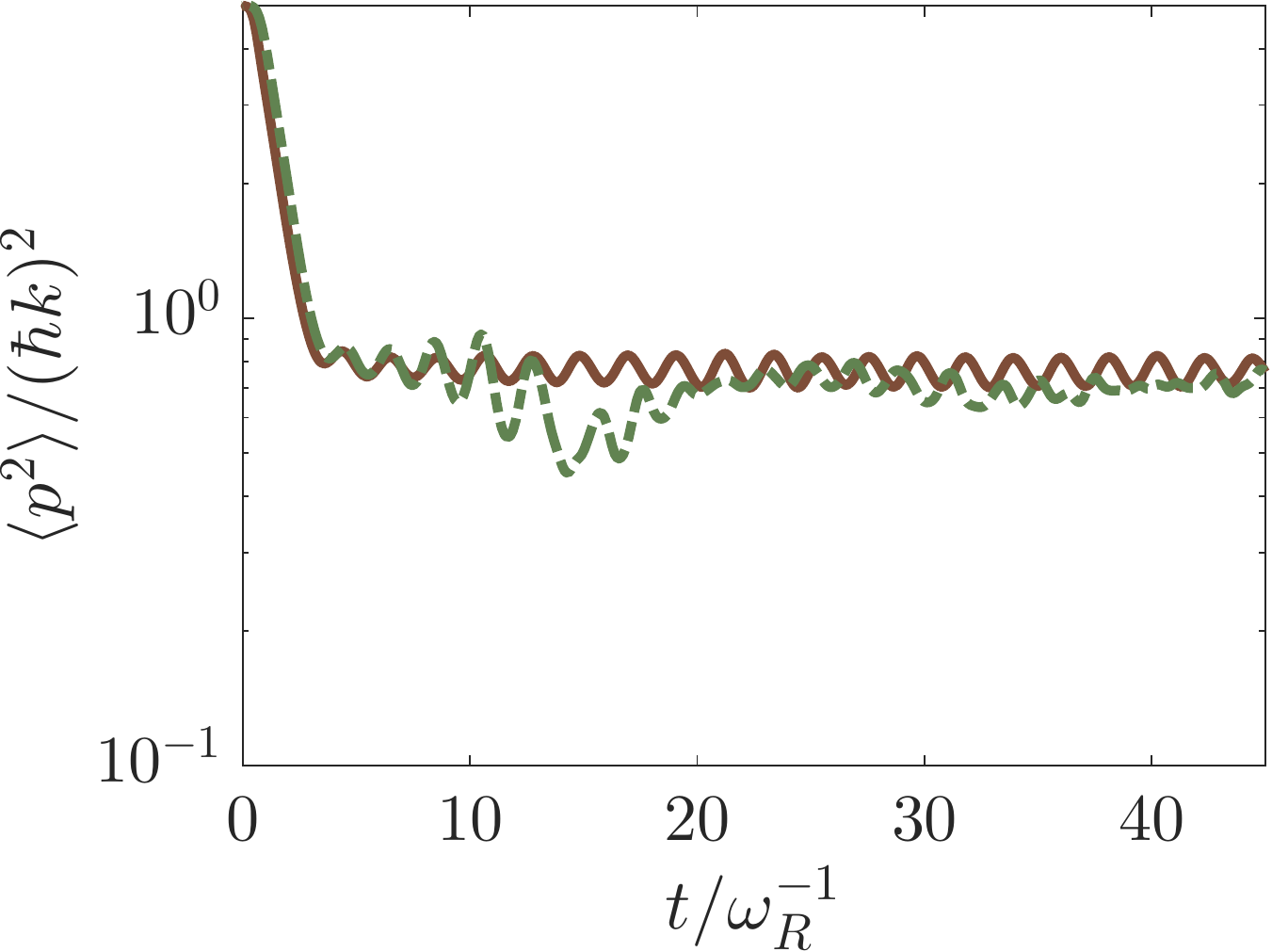}
	\caption{Dynamics of $\langle p^2(t)\rangle$ (in units of $\hbar^2 k^2$) as a function of time (in units of $\omega_R^{-1}$) for (a) $N=100$ and (b) $N=1000$ atoms. The solid lines are obtained by numerically integrating Eqs. \eqref{sigma2},\eqref{sigmaz2} (using the cumulant expansion), and \eqref{scfull}, whereas the dashed lines are the predictions of the mean field model in Eqs. \eqref{sj},\eqref{zj},\eqref{meanfieldforce} . The other parameters are the same as in Fig. \eqref{tempcomparison}(a). Note that $N\Gamma_C=40\omega_R$. Accordingly, we rescale the value of $\Gamma_C$ when increasing $N$.}
	\label{N100mean}
\end{figure}

\begin{figure}[h!]
	\flushleft(a)\vspace{-4ex}\\
\center\includegraphics[width=0.8\linewidth]{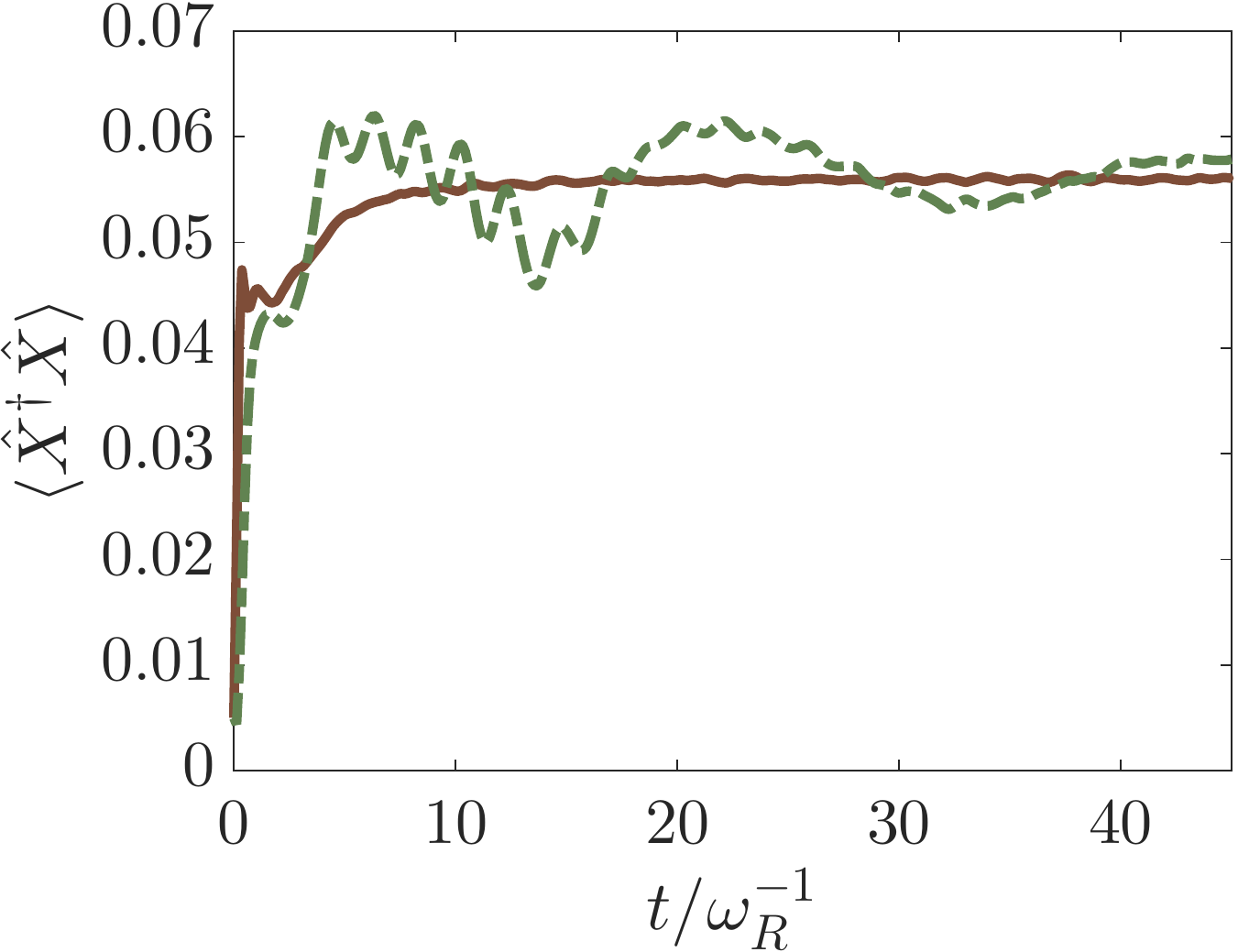}
	\flushleft(b)\vspace{-4ex}\\
	\center\includegraphics[width=0.8\linewidth]{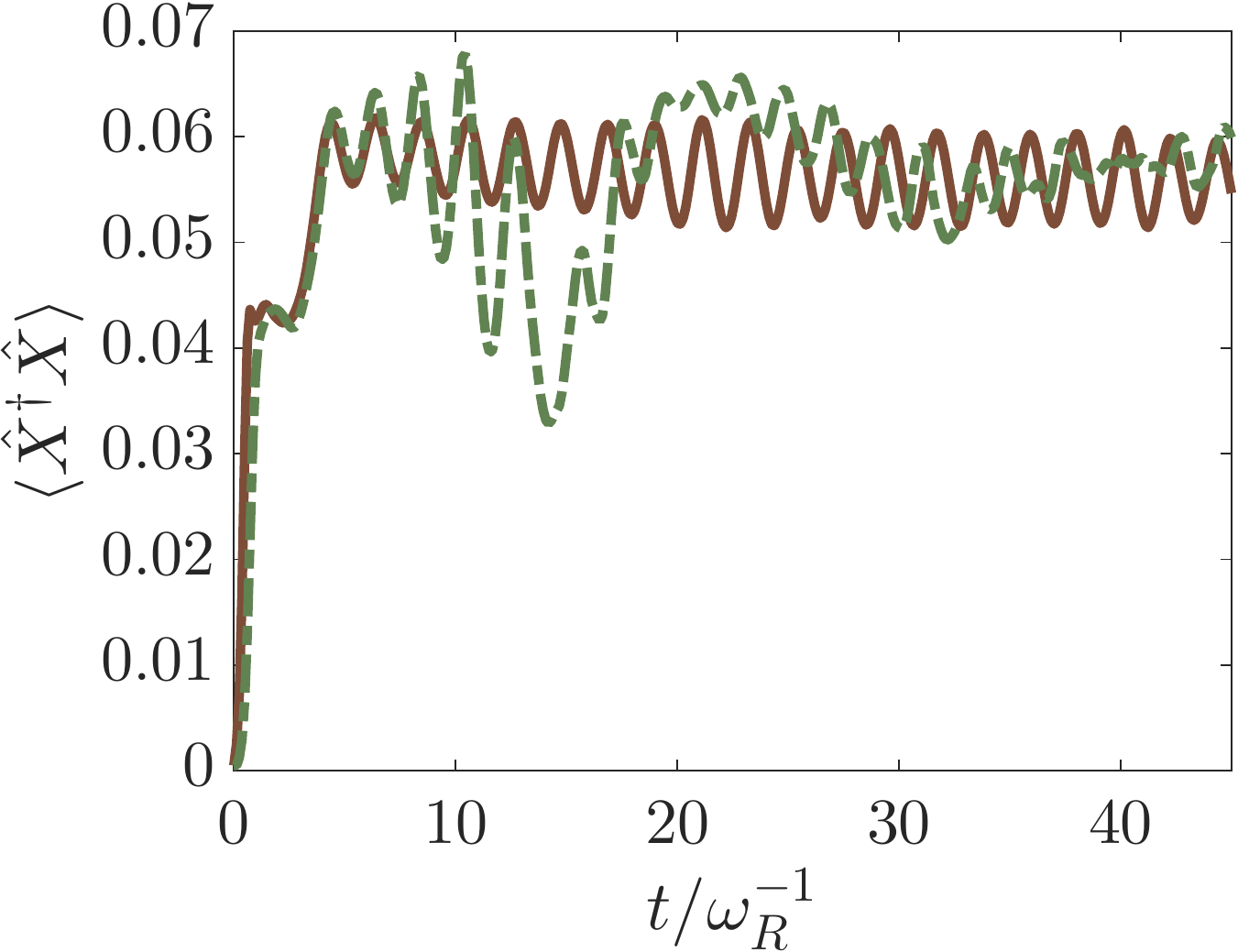}
	\caption{Dynamics of $\langle \hat{X}^{\dag}\hat{X}\rangle$ as a function of time (in units of $\omega_R^{-1}$ for (a) $N=100$ and (b) $N=1000$, corresponding to the subplots of Fig. \ref{N100mean}.}
	\label{N1000mean}
\end{figure}

Figure  \ref{N100mean} displays the dynamics of $\langle p^2(t)\rangle$ predicted by the semiclassical model (solid line) and by the mean-field model (dashed lines) for $\langle p^2(0)\rangle=5\,(\hbar k)^2$. The two curves qualitatively agree. Moreover, their behaviour at short times almost coincides and the time interval over which this occurs increases with $N$. A striking difference is the small frequency oscillation, which seems to solely characterize the mean-field dynamics. However, this oscillation becomes visible at time scales at which the mean-field and the semiclassical dynamics start to be quantitatively distinct. The fast oscillations, instead, are also reproduced by the semiclassical equations at $N=1000$. They are also visible in the dynamics of the expectation value of $\langle\hat{X}^{\dag}\hat{X}\rangle$, as shown in Fig. \ref{N1000mean}. We note that the mean-field and full semiclassical dynamics predict approximately the same stationary values of the correlations.

\begin{figure}[h]
	\center \includegraphics[width=0.8\linewidth]{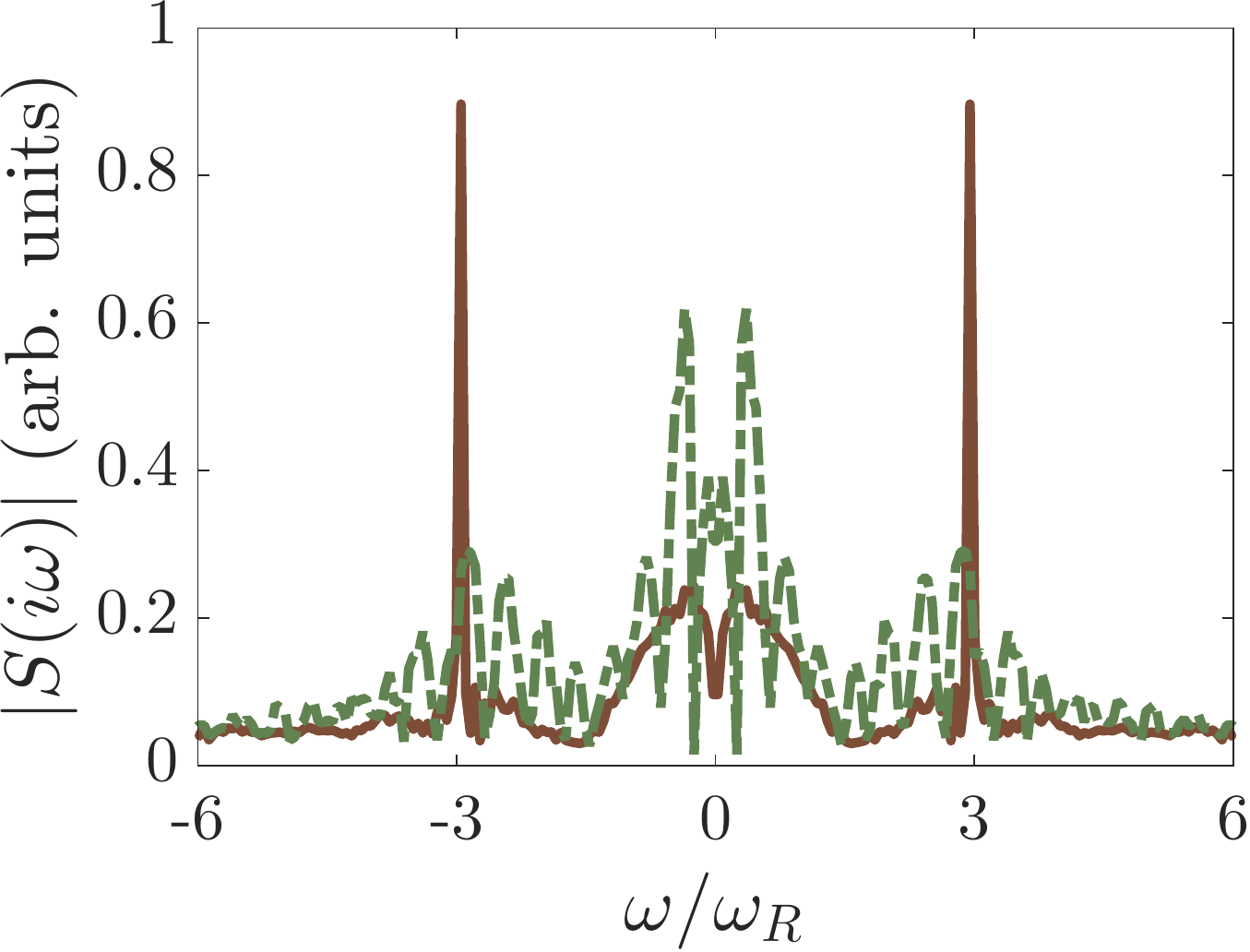}
	\caption{Absolute value of the Laplace transform $S(i\omega)$, Eq. \eqref{S:w}, in arbitrary units and as a function of $\omega$ (in units of $\omega_R$) for the curves in Fig. \ref{N1000mean}(b). The Laplace transform is evaluated over the same time interval as in Fig. \ref{N1000mean}(b).}
	\label{fourierplot}
\end{figure}

Figure  \ref{fourierplot} displays the spectral analysis of the two curves in Fig. \ref{N1000mean}(b). In detail, it illustrates the Laplace transform $S(i\omega)$, defined as: 
\begin{equation}
\label{S:w}
S(i\omega)=\int_{0}^{\infty}e^{i\omega t}\left(\langle\hat{X}^{\dag}\hat{X}\rangle(t)-\langle\hat{X}^{\dag}\hat{X}\rangle_{\text{st}}\right)dt\,,
\end{equation}
where $\langle\hat{X}^{\dag}\hat{X}\rangle_{\text{st}}=\lim_{t\to\infty}\langle\hat{X}^{\dag}\hat{X}\rangle(t)$. The spectrum of the mean-field data (dashed-dotted curve) and of the data predicted by the semiclassical model (solid curve) exhibits two sidebands at $\omega\simeq \pm 3\omega_R$, which we attribute to the oscillations in the potential confining the atoms (see next section). The mean-field simulations predict also two low frequency sidebands at a frequency of the order of a fraction of the recoil frequency, which correspond to the slow oscillations observed in Fig. \ref{N1000mean}(b).  

\subsection{Dynamics at the asymptotic limit}

Using the mean-field model we now investigate the dynamics at the asymptotic limit. In particular, we assume that the atoms are sufficiently cold so that at this stage $T_i\gg T_e$. It is therefore justified to adiabatically eliminate the internal degrees of freedom from the equations of motion of the atoms' external variables. The procedure is detailed in Appendix \ref{App:C} and leads to the stationary values $s_j^{(0)}$ and $z_j^{(0)}$, which read
\begin{align}
s_j^{(0)}=&\frac{\xi(x_j)}{1+2|\xi(x_j)|^2},\label{eliminatedsj}\\
z_j^{(0)}=&\frac{1}{1+2|\xi(x_j)|^2}\label{eliminatedzj},
\end{align}
and thus depend on the atomic position $x_j$ through the quantity
\begin{equation}
\xi(x_j)=\frac{N\Gamma_C}{w}X\cos(kx_j)\,.
\end{equation}
This quantity is proportional to the ratio $N\Gamma_C\cos(kx_j)/w$. It plays an analogous role to the saturation parameter in the dynamics of a driven dipole \cite{CohenTannoudji:Book}, but its source is of a completely different nature: it depends on the synchronization order parameter $X$, which is found by solving self-consistently the equation
\begin{align}
\frac{1}{N}\sum_{j=1}^N\frac{|\xi(x_j)|^2}{1+2|\xi(x_j)|^2}=\frac{N\Gamma_C}{w}|X|^2\,.\label{Xmean}
\end{align}
Concise solutions, which are limiting cases, can be found by assuming that the atoms are tightly confined in a lattice, thereby fixing $\cos (k x_j)=\pm\delta$. For $\delta=0$, for instance, the only solution is $X=0$. For $\delta\neq 0$, instead, one finds
\begin{align}
|X|^2=\frac{w}{2N\Gamma_C}\left(1-\frac{w}{\delta^2N\Gamma_C}\right)\,.
\end{align}
From this equation it follows that $|X|^2=0$ both when $w=0$ and also when $w\geq N\Gamma_C \delta^2$, namely when $w$ takes the value of the upper synchronization threshold for the corresponding configuration. In particular, the upper synchronization threshold is maximum when $\delta=1$, which corresponds to the value reported in Ref. \cite{Meiser:2010:1}. 

When instead the atoms are uniformly distributed over the cavity wavelength, Eq. \eqref{Xmean} can be recast in the form 
\begin{align}
\frac{N\Gamma_C}{w}\frac{1}{2\pi}\int_0^{2\pi}d\theta \frac{\cos^2\theta}{1+2(N\Gamma_C/w)^2|X|^2\cos^2\theta}=1\,.\label{Xmean:2}
\end{align}
By using 
\begin{align*}
&\frac{1}{2\pi}\int_0^{2\pi}d\theta \frac{\cos^2\theta}{1+2(N\Gamma_C/w)^2|X|^2\cos^2\theta}\\
&=\frac{1}{1+2(N\Gamma_C/w)^2|X|^2+\sqrt{1+2(N\Gamma_C/w)^2|X|^2}}\,,
\end{align*}
we get
\begin{align}
|X|^2=\frac{w}{2N\Gamma_C}\left(1-\frac{w}{N\Gamma_C}\left(\frac{1}{2}+\sqrt{\frac{N\Gamma_C}{w}+\frac{1}{4}}\right)\right)\label{Xmean:3}\,.
\end{align}
From Eq. \eqref{Xmean:3} one obtains that the upper synchronization threshold for particles that are homogeneously distributed over the cavity wavelength is given by $w=N\Gamma_C/2$.\\

\begin{figure}[h]
	\flushleft(a)\vspace{-4ex}\\
	\center \includegraphics[width=0.8\linewidth]{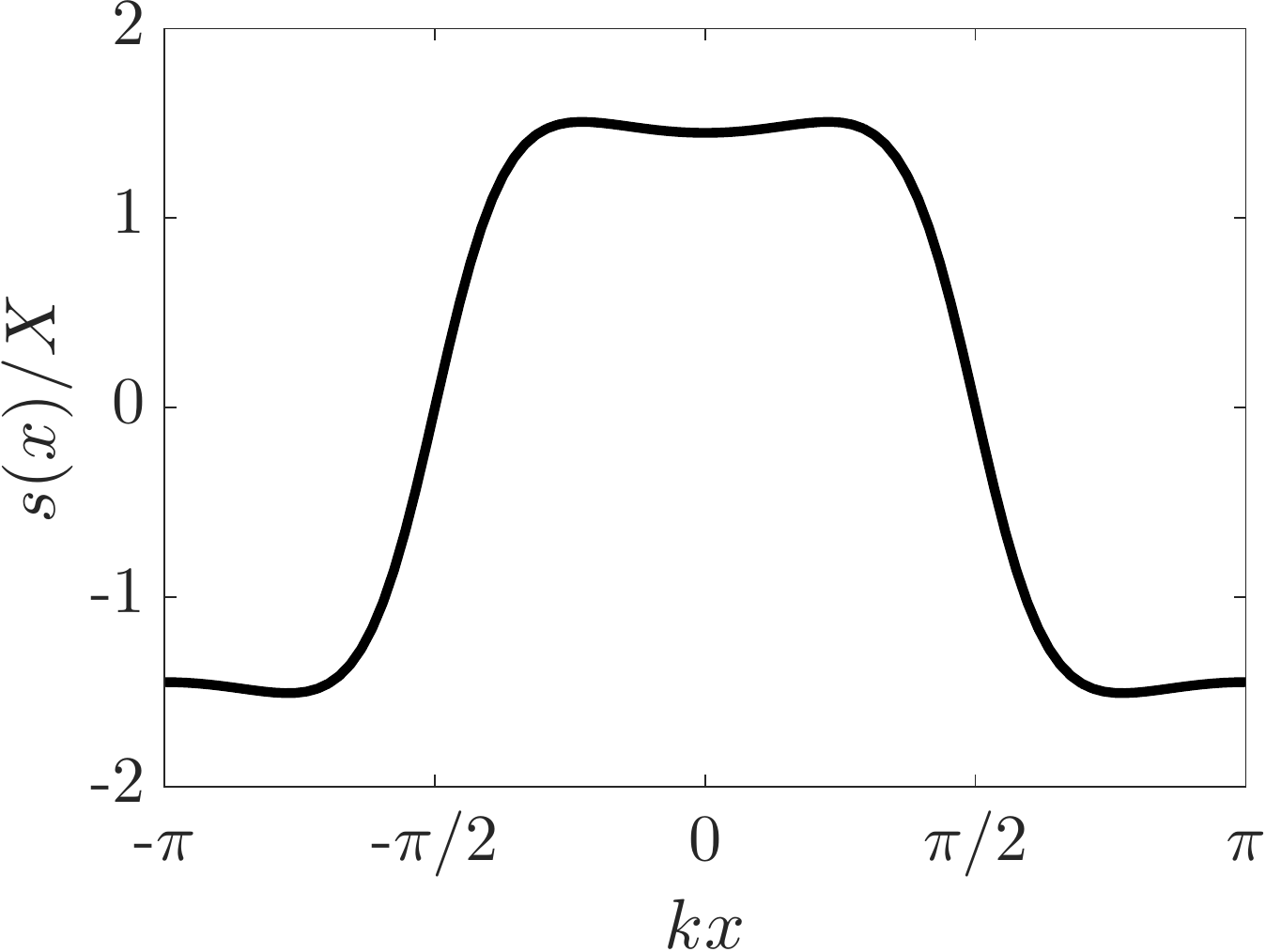}
	\flushleft(b)\vspace{-4ex}\\
	\center\includegraphics[width=0.8\linewidth]{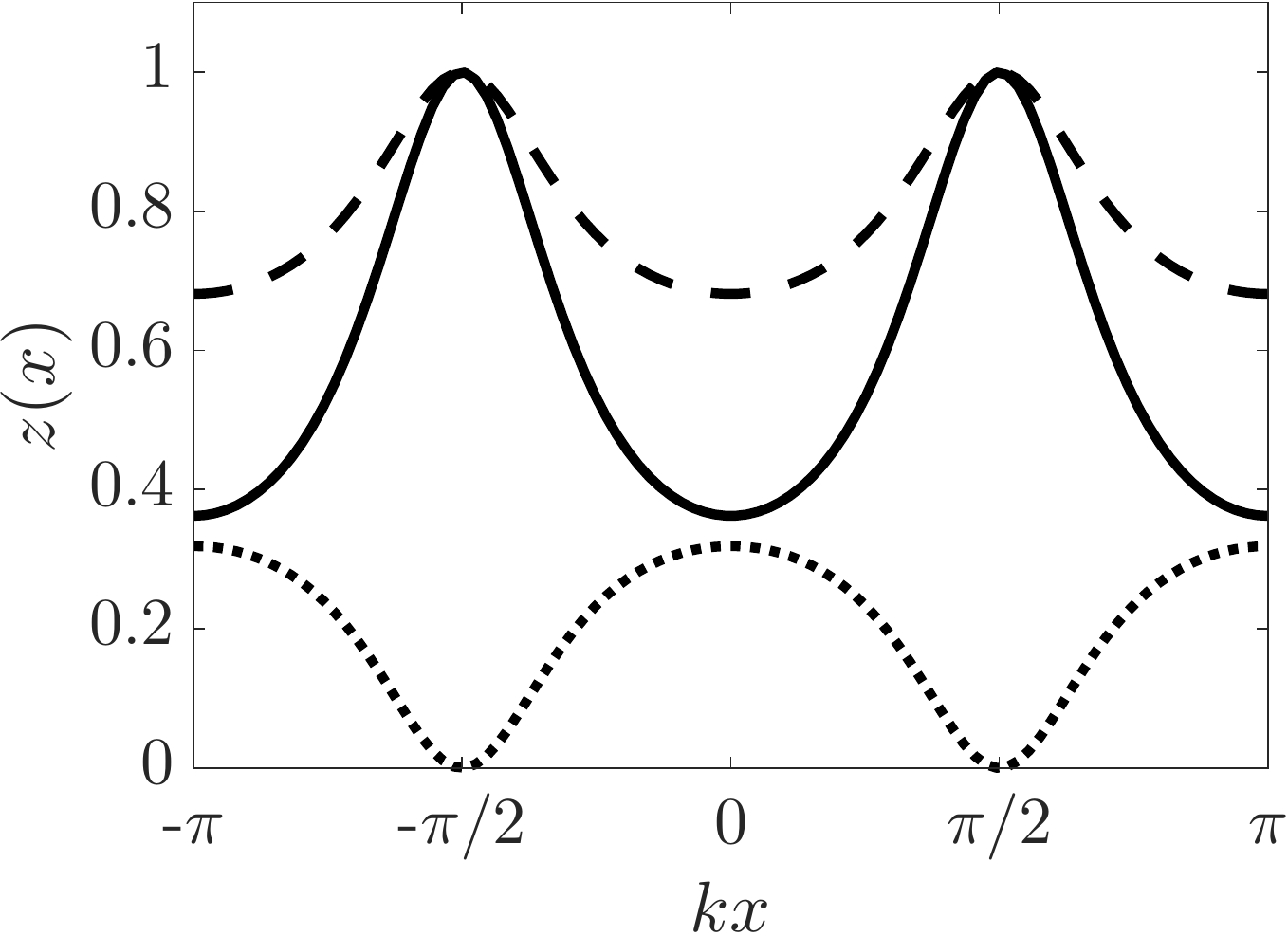}
		\caption{Spatial dependence of (a) the dipole moment $s(x)$ (Eq. \eqref{eliminatedsj}) and (b) the population inversion $z(x)$ (Eq. \eqref{eliminatedzj}) for $w=N\Gamma_C/4$ and $|X|^2\approx0.055$. The $x$-axis is in units of $1/k$. We verified that this behaviour is also predicted by the semiclassical model.}
		\label{sz}
\end{figure}

We now use this result to determine the spatial dependence of the dipole moment $s_j^{(0)}$ and of the population inversion $z_j^{(0)}$. These two quantities are plotted in Fig. \ref{sz}(a) and (b), respectively, where we have used the definition $s_j^{(0)}\to s(x)$ ($z_j^{(0)}\to z(x)$) in the continuum limit. We observe that at the nodes  of $\cos(kx)$ the polarization $s(x)$ changes its sign where the population inversion is maximum. In turn, the population inversion is  minimal close to the antinode where the polarization reaches its maximum absolute value. If one associates a well-defined magnetic moment to the two electronic states, then the resulting behaviour corresponds to an effective anti-ferromagnetic order. 

\subsection{Effective Hamiltonian}

In the adiabatic limit, where one neglects retardation effects in the coupled dynamics between spin and center-of-mass motion, it is possible to derive an effective Hamiltonian for the atomic external variables. For this purpose we use Eqs. \eqref{eliminatedsj} and \eqref{eliminatedzj} in Eq. \eqref{meanfieldforce} to obtain:
\begin{align*}
\dot{p}_j=-\hbar kw\frac{\Delta}{\kappa/2}\tan(kx_j)\frac{|\xi(x_j)|^2}{1+2|\xi(x_j)|^2}\,.
\end{align*}
For $N\gg 1$ we can write this equations as $\dot p_j=-\partial V_{\mathrm{eff}}/\partial x_j$ where $V_{\mathrm{eff}}$ is an effective potential of the form:
\begin{align}
V_{\mathrm{eff}}=-\sum_{j=1}^N\frac{\hbar w}{4}\frac{\Delta}{\kappa/2}\log\left(1+2|\xi(x_j)|^2\right) \,.\label{Veff}
\end{align}
The corresponding mean-field Hamiltonian, $H_{\mathrm{mean}}$, reads
\begin{align}
H_{\mathrm{mean}}=\frac{p^2}{2m}-\frac{\hbar w}{4}\frac{\Delta}{\kappa/2}\log\left(1+2|\xi(x)|^2\right).\label{meanHamiltonian}
\end{align}
The potential minima are at the positions $x$ where $\cos(kx)=\pm1$.
At these points the atoms would be trapped should their asymptotic temperature be smaller than the potential depth $-\hbar\Delta w/(2\kappa)\log(1+2N\Gamma_C|X|_{1}^2/w)$ (here $|X|_{1}$ is the synchronization order parameter when the atoms are confined at $\cos(kx)=\pm 1$). Correspondingly, the atoms would form an antiferromagnetic spin chain, where the spins swap their orientation so to keep $s_j\cos(kx_j)=1$. In order to verify whether this is the stationary state of the synchronization dynamics, one needs first to determine the asymptotic temperature. Part of this analysis is performed in the next section, where we determine the friction force due to the non-adiabatic coupling with the internal degrees of freedom. 

\subsection{Dissipative mean-field dynamics}

Retardation effects in the dynamics of the spins following the motion give rise to friction. The steady state results from the interplay between the friction and the dispersive force due to the effective potential. 
We now determine the friction forces in the last cooling stage. For this purpose we perform an expansion of the spin variables including terms to first order in the small parameter ${\mathcal R}_{\rm Doppler}/w$:
\begin{align*}
s=&s^{(0)}+\frac{kp}{mw}s^{(1)}\,,\\
z=&z^{(0)}+\frac{kp}{mw}z^{(1)}\,.
\end{align*}
We then use the prescription $d/dt\to\partial/\partial t+p/m\partial/\partial x$ in Eqs. \eqref{sj} and \eqref{zj} and determine the corresponding stationary state (see Appendix \ref{App:D} for details). 

The friction force is the component of the force in Eq. \eqref{meanfieldforce} that depends on the retarded component, $$F_{\mathrm{ret}}=-(\hbar k/m)p\sin(kx)N\Gamma_C\mathrm{Re}\left\{\alpha X^*s^{(1)}\right\}\,,$$ and takes the form
	\begin{eqnarray}
	&&F_{\mathrm{ret}}=-8\omega_R p\frac{|\xi(x)|^2\tan^2(kx)}{(1+2|\xi(x)|^2)^3}\frac{\Delta}{\kappa/2}\mathcal F_{\Delta}(|\xi(x)|^2)
	\label{friction}
	\end{eqnarray}
	with
	\begin{eqnarray}
	&&\mathcal F_{\Delta}(|\xi(x)|^2)=\frac{{1-2|\xi(x)|^2}}{1+\left(\frac{\Delta}{\kappa/2}\right)^2}-2|\xi(x)|^4\,.\nonumber
	\end{eqnarray}
This equation shows that the friction force depends on the atomic position. It vanishes at the minima of the mean-field potential, where $\sin(kx)=0$, but tends to pull out the atoms from these points, being positive about $x=n\pi/k$ for $\Delta>0$. The friction coefficient is instead negative for values of $\zeta(x)$ such that
\begin{equation}
|\xi(x)|^2\le \frac{1}{2|\alpha|^2}\left(\sqrt{2|\alpha|^2+1}-1\right)\,.
\end{equation}
The equality holds at the positions $x_0$, where the force changes sign. Hence, at the positions $x$ where $\cos^2(kx)<\cos^2(kx_0)$ the friction force is negative. Remarkably, these positions are close to the maxima of the mean-field potential. 
\begin{figure}[h]
	\flushleft(a)\vspace{-4ex}\\
	\center \includegraphics[width=0.8\linewidth]{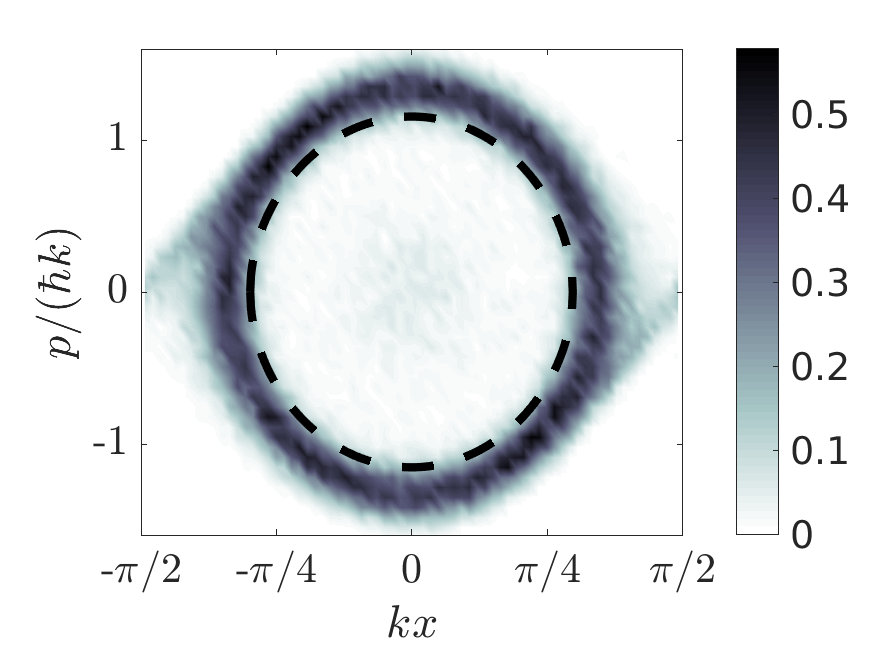}
	\flushleft(b)\vspace{-4ex}\\
	\center\includegraphics[width=0.8\linewidth]{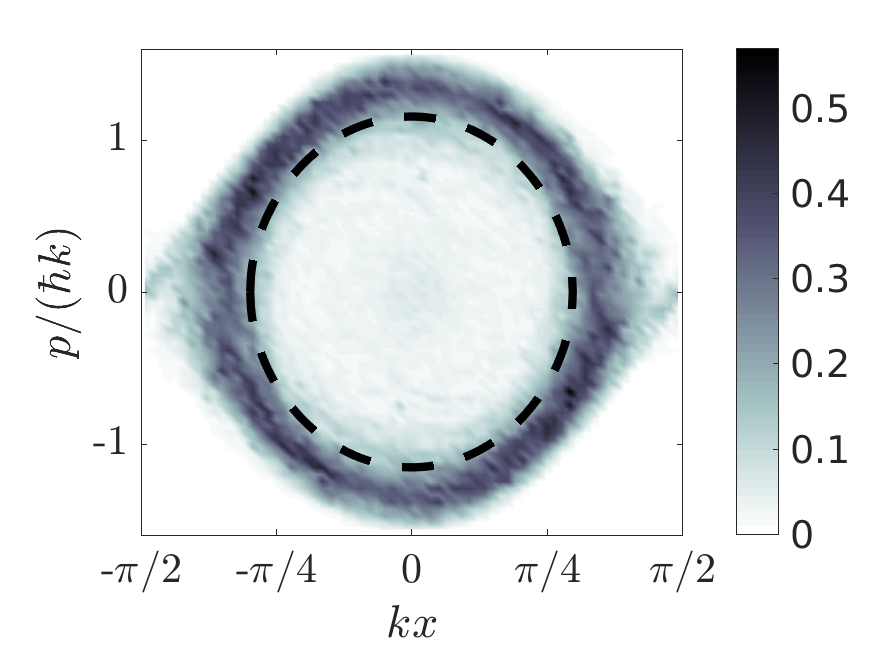}
	\caption{Phase space histogram of the asymptotic dynamics of $N=1000$ particles, the $x$ axis is in unit of $1/k$ and the trajectories are reported modulus the wavelength; the $p$ axis is in units of $\hbar k$. The parameters are the same as in Fig. \ref{N100mean}(b), the time is of the order of $t\approx 50\omega_R^{-1}$. Subplot (a) reports $100$ trajectories calculated using stochastic differential equations \cite{SDE} simulating the dynamics of Eqs. \eqref{sigma2},\eqref{sigmaz2}, and  \eqref{scfull}. Subplot (b) reports the corresponding mean-field simulations of Eqs. \eqref{sj},\eqref{zj},\eqref{meanfieldforce}. The black dashed line indicates the trajectory at energy $E_0$, Eq. \eqref{E0}. }
	\label{phasespaceplots}
\end{figure}

\begin{figure}[h]
	\flushleft(a)\vspace{-4ex}\\
	\center \includegraphics[width=0.8\linewidth]{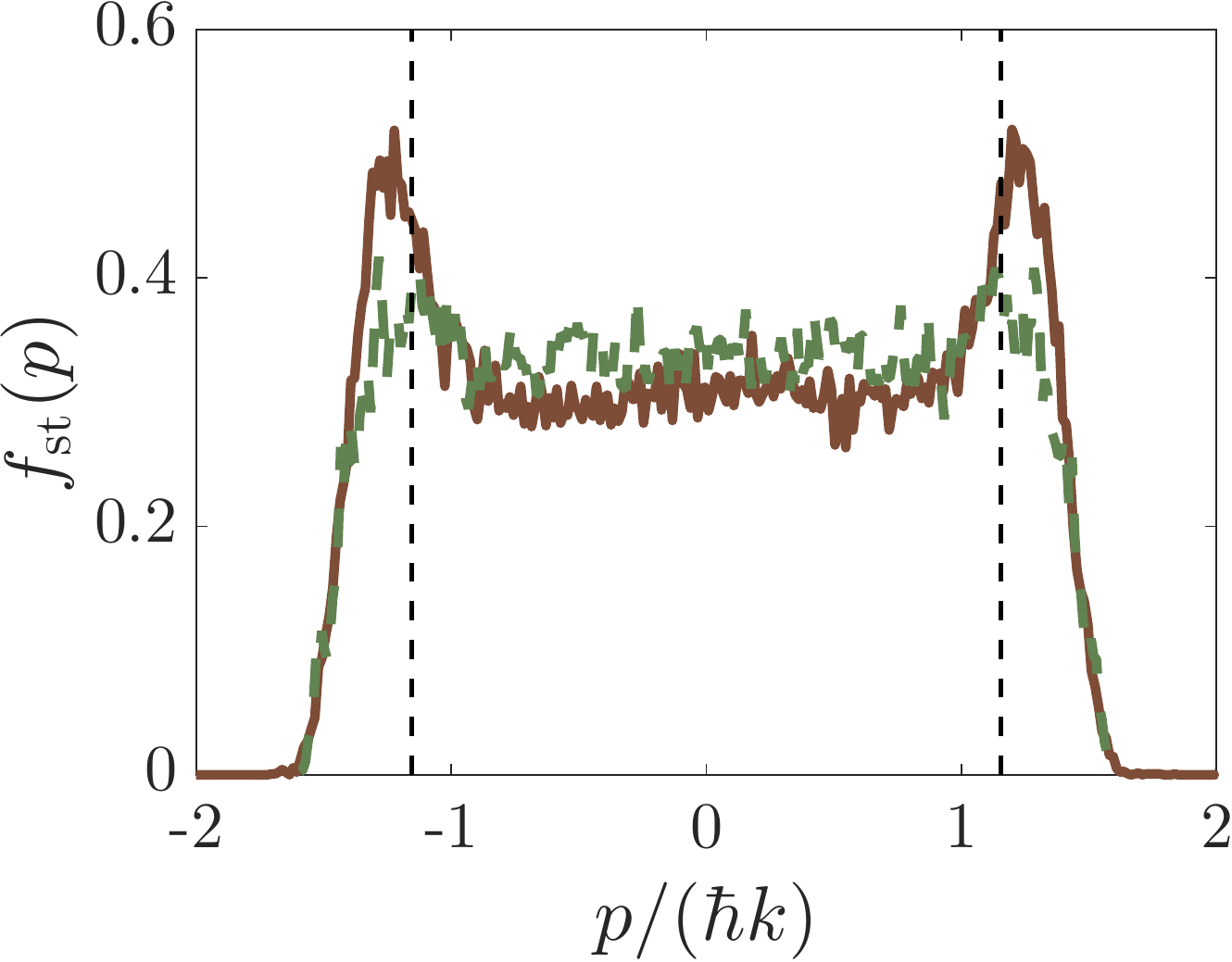}
	\flushleft(b)\vspace{-4ex}\\
	\center\includegraphics[width=0.8\linewidth]{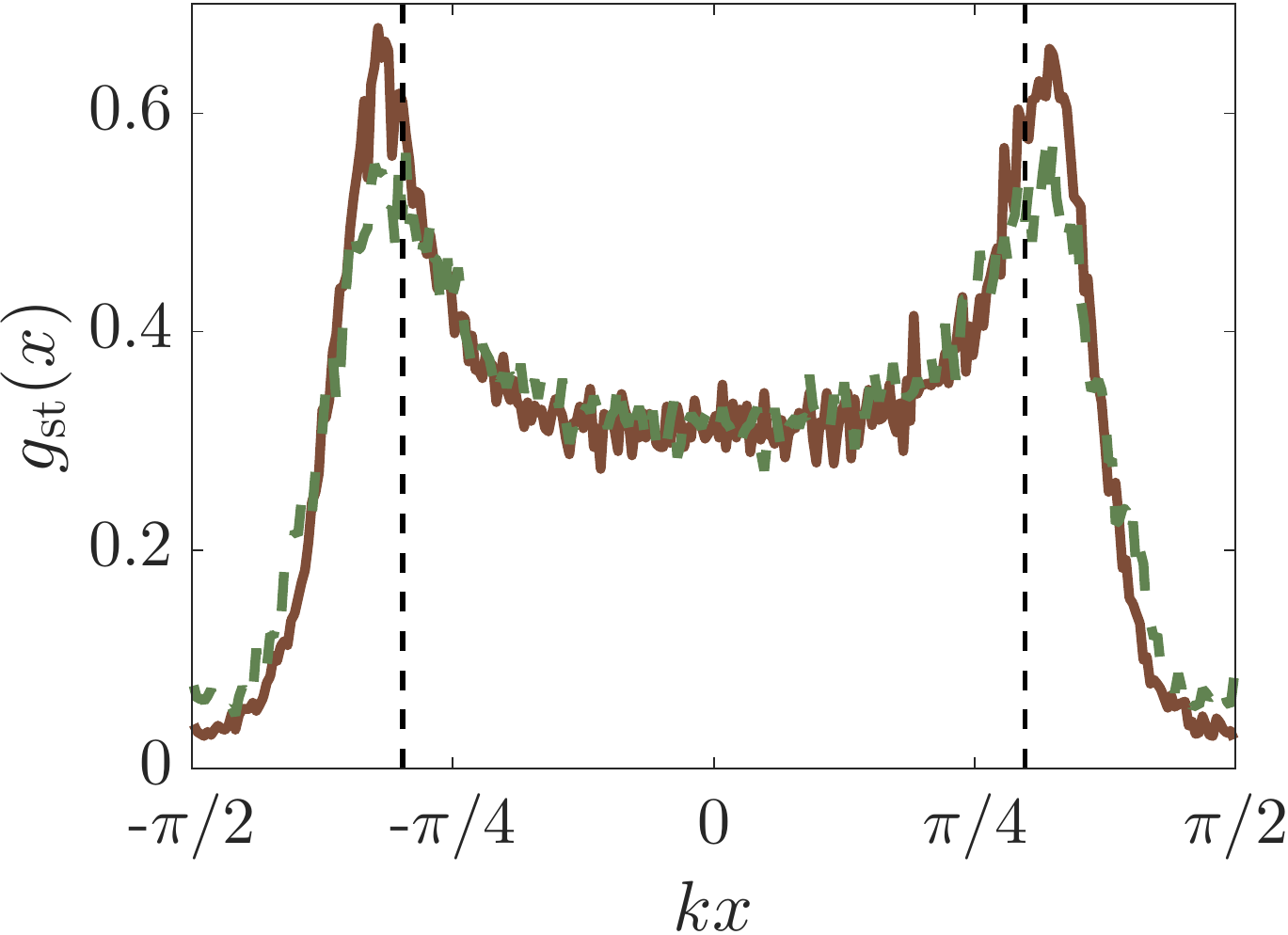}
	\caption{(a) Momentum distribution $f_{\text{st}}(p)$ as a function of $p$ (in units of $\hbar k$) and  position distribution $g_{\text{st}}(x)$ as a function of $x$ (in units of $1/k$ and modulus the wavelength) corresponding to the distribution in Fig. \ref{phasespaceplots} . The brown solid line shows the prediction of the semiclassical simulation (corresponding to Fig. \ref{phasespaceplots}(a)),  the green dashed line shows the prediction of the mean-field model (Fig. \ref{phasespaceplots}(b)).}
	\label{distributions}
\end{figure}

Figure \ref{phasespaceplots}(a) displays  the trajectories in phase space at steady state obtained by integrating the semiclassical equations, while subplot (b) shows the corresponding prediction of the mean-field model. 
Comparison between subplots (a) and (b) shows that the resulting trajectories form rings centered at $p=0$ and at the points $x=n\pi$ with $n$ denoting any integer number. The rings are connected and the trajectories are indeed close to the separatrix. The separatrix represents the separation of the trajectories where the atoms are bound at the mean-field potential minima from the trajectories where the atoms are unbound. The dashed line indicates in particular the trajectory where the kinetic energy vanishes at the roots $x_0$ (namely, where the non-conservative force changes sign). Its energy is given by
\begin{align}
E_0=&-\frac{\hbar w}{4}\log\left(1+2|\xi(x_0)|^2\right) \label{E0}\,.
\end{align}
A careful comparison between subplots (a) and (b) shows that cavity shot noise (included in the simulation of (a)) tends to suppress the trajectories with energy below $E_0$. Figures  \ref{distributions} (a) and (b) report the corresponding momentum and position distributions, respectively. The momentum distribution, Fig. \ref{distributions} (a), is almost flat over the interval $[-p_0,p_0]$, such that $p_0^2\sim 2mE_0$ (these points are indicated by the vertical dashed lines).  The semiclassical simulations predict at these specific points two peaks, which are otherwise absent in the mean-field prediction. Instead, mean field and semiclassical simulations deliver very similar position distributions,  as shown in Fig. \ref{distributions} (b). Here, the two peaks of the distribution are located about the positions $x_0$ where the non-conservative force vanishes.\\

\subsection{Asymptotic temperature}

\begin{figure}[h]
	\flushleft(a)\vspace{-4ex}\\
	\center \includegraphics[width=0.75\linewidth]{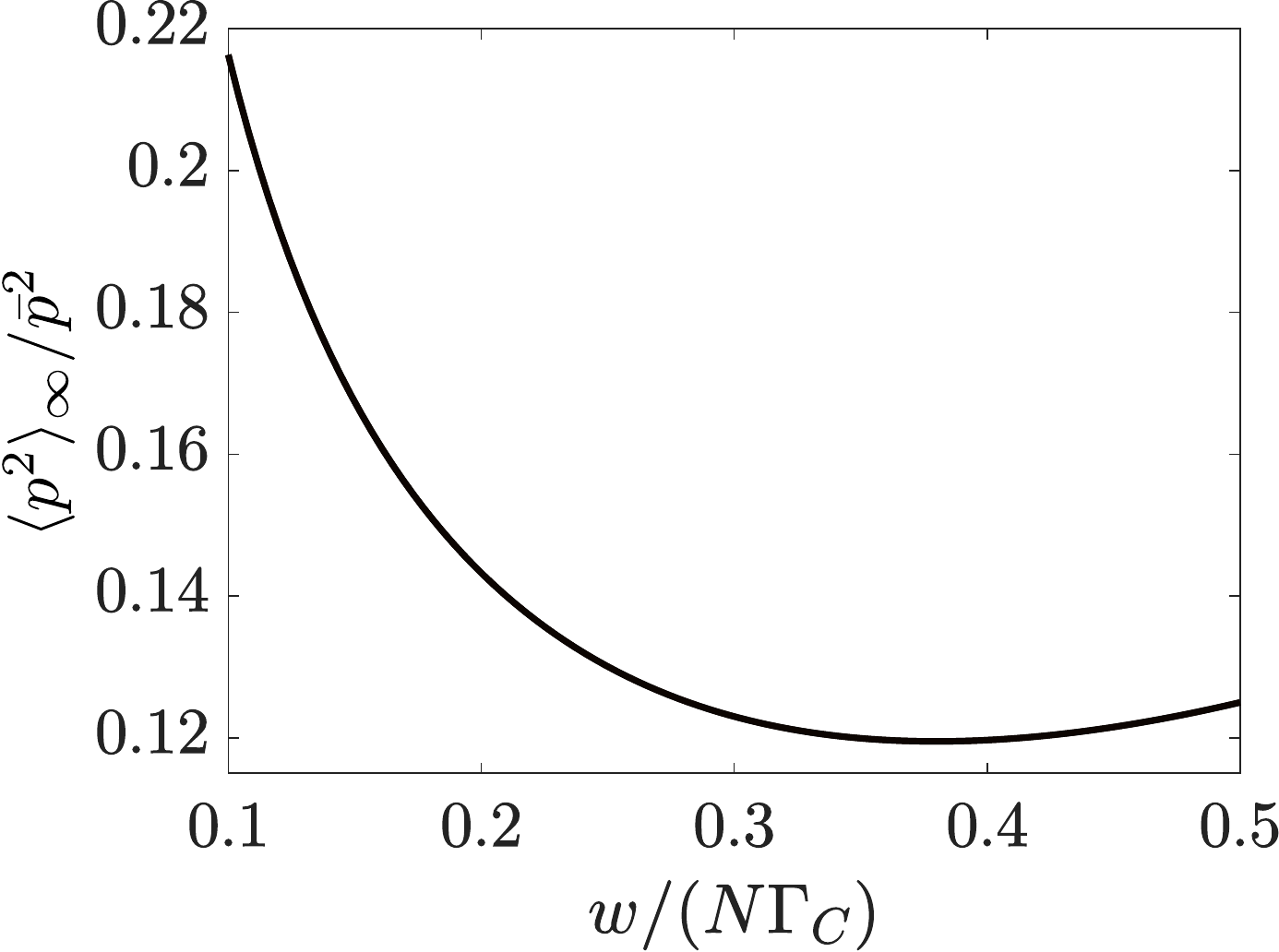}\\
	\flushleft(b)\vspace{-4ex}\\
	\center \includegraphics[width=0.75\linewidth]{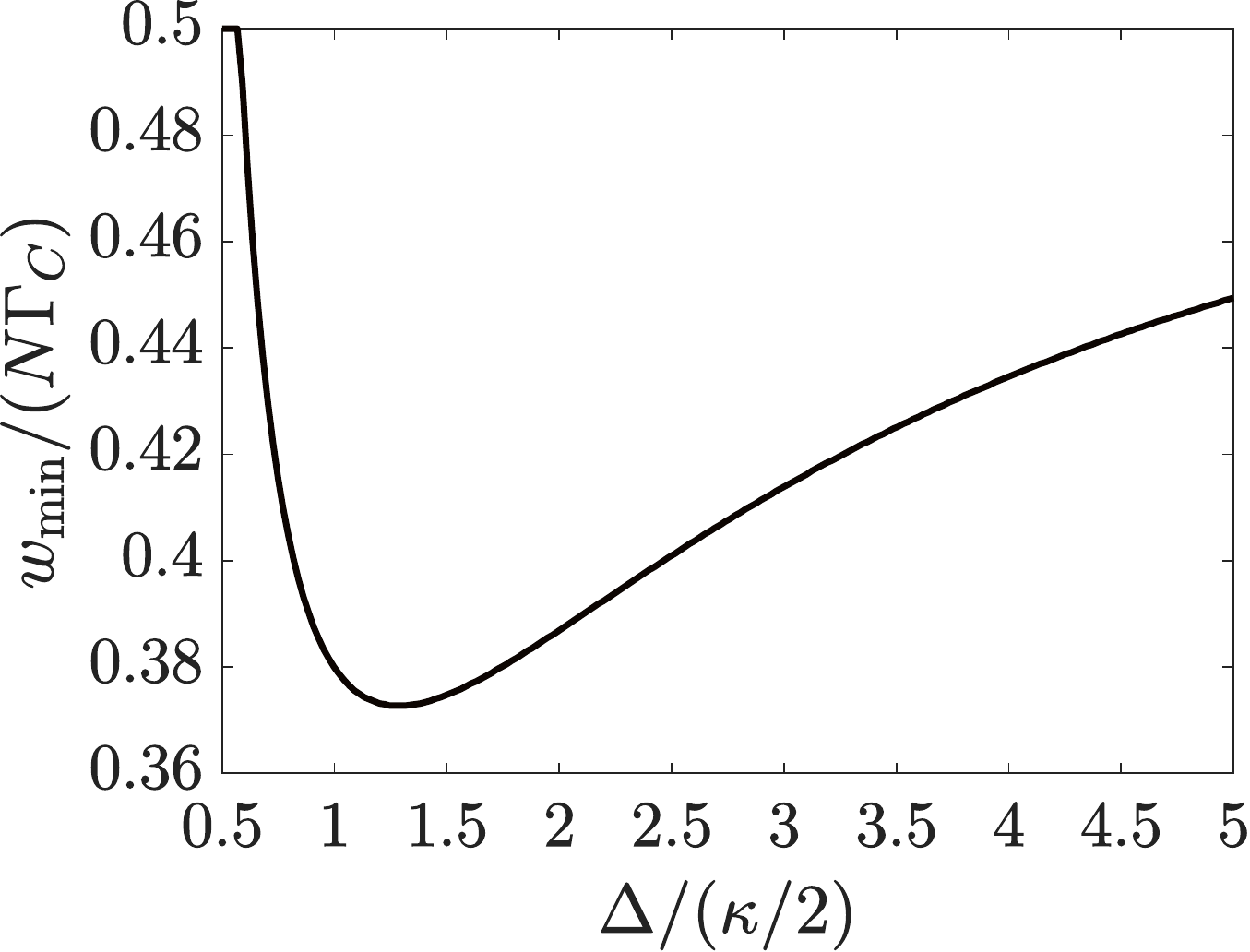}\\
	\flushleft(c)\vspace{-4ex}\\
	\center \includegraphics[width=0.75\linewidth]{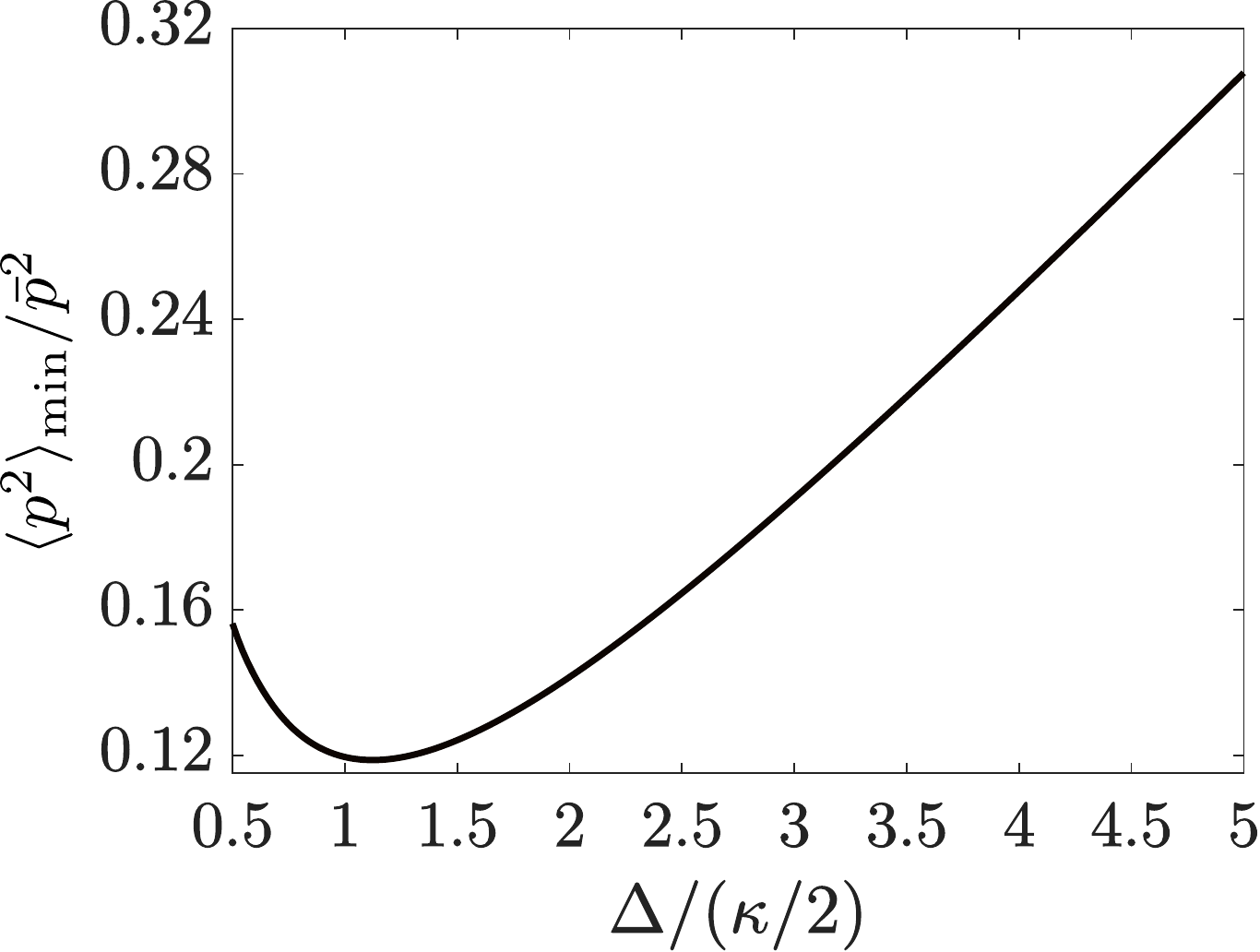}\\
	\caption{Stationary momentum width $\langle p^2\rangle_\infty$ (in units of $\bar{p}^2=(\hbar k)^2N\Gamma_C/(2\omega_R)$) as a function of $w$ (in units of $N\Gamma_C$) and for $\Delta=\kappa/2$. Subplot (b) shows the value of the pumping strength $w_{\mathrm{min}}$  (in units of $N\Gamma_C$)  that minimizes the temperature for each value of the detuning $\Delta$ (in units of $\kappa/2$). Subplot (c) reports the corresponding value of the minimum width $\langle p^2\rangle_{\mathrm{min}}$ (in units of $\bar{p}^2$) as a function of $\Delta$  (in units of $\kappa/2$).}
	\label{temperature}
\end{figure}

We now estimate the asymptotic temperature by means of the fluctuation-dissipation theorem. The validity of the theorem is limited, since the stationary momentum distribution is not thermal, but allows us to gain insight into the dependence of the momentum distribution on the physical parameters. In what follows we extract the friction coefficient $\gamma$ from the force in Eq. \eqref{friction}:
\begin{align}
	\gamma(x)=8\omega_R\frac{|\xi(x)|^2\tan^2(kx)}{(1+2|\xi(x)|^2)^3}\frac{\Delta}{\kappa/2}\mathcal F_{\Delta}(|\xi(x)|^2)\,.
	\end{align}
The calculations for the diffusion coefficients include spin noise due to the incoherent pump, their derivation is involved and reported in Appendix \ref{App:E}. The resulting diffusion coefficient $D(x)$ is given in Eq. \eqref{Diffusioncoefficient}. The final width of the momentum distribution is found after integrating $D(x)$ and $\gamma(x)$ over the asymptotic atomic spatial density distribution, which for convenience we assume to be uniform. Denoting by $\bar D$ and $\bar \gamma$ the corresponding average values, we obtain 
\begin{align}
\langle p^2\rangle_\infty=\frac{\bar D}{\bar \gamma}\,, \label{finaltemp}
\end{align}
where $\langle p^2\rangle_\infty=\lim_{t\to\infty}\langle p^2(t)\rangle$.
Figure \ref{temperature}(a) displays the ratio of Eq. \eqref{finaltemp} as a function of the pump rate $w$ and for $\Delta=\kappa/2$. The minimal width is reached at a value between $w=N\Gamma_C/10$ and $w=N\Gamma_C/2$. For  $w=N\Gamma_C/2$, in particular,  $$\frac{\langle p^2\rangle_{\infty}}{2m}=\frac{\hbar w}{8}=\frac{\hbar N\Gamma_C}{16}\,.$$
Figure \ref{temperature}(b) shows the value of the pump rate which minimizes the momentum width as a function of $\Delta$. The corresponding temperature is shown as a function of $\Delta$ in subplot (c) and is minimized at $\Delta\approx\kappa/2$. The results suggest that lower temperatures can be reached by decreasing $N\Gamma_C$ (as long as this value is larger than the recoil frequency, consistently with the semiclassical treatment here applied). 

\subsection{Discussion}

The setup we analyse in this work is the same as the one discussed in Ref. \cite{Salzburger:2006}, nevertheless the studies in those papers focus on different parameter regimes which lead to substantially different dynamics.
These works predict lasing as well as spatial localization of the atoms in steady state when the atoms are incoherently pumped from the side. The model of Ref.  \cite{Salzburger:2006}, in particular, focuses on the dynamics of an atomic ensemble. It includes spontaneous emission and assumes that the rate of the incoherent pump is the largest parameter of the dynamics. With this choice population inversion is achieved. 

A key point is that the faster time scale of the dynamics in Ref. \cite{Salzburger:2004,Salzburger:2005,Salzburger:2006} is determined by the pump rate and spontaneous decay. For this reason the regime is reached where the atomic internal degrees of freedom follow adiabatically the coupled dynamics of the external and of the resonator degrees of freedom. This is warranted when the atoms are localized in the antinodes of the cavity standing wave. This choice justifies the approximation of discarding terms such as $\langle \hat{\sigma}^{\dag}_j\cos(kx_j)\hat{\sigma}_i\cos(kx_i)\rangle$ for $i\neq j$ in their equations (which also means that this regime cannot exhibit synchronization). This assumption leads to the scaling of the intracavity photon number with the number of atoms in the excited state and thus with $N$ (see Appendix \ref{App:F} for further details). 

In contrast, here we consider the regime in which the cavity decay $\kappa$ sets the fastest time scale. Moreover, we choose the values of the pump rate for which synchronization is expected. As a result, the dynamics we predict is intrinsically due to collective effects, since it is dominated by mean-field correlations $\langle \hat{X}^{\dag}\hat{X}\rangle\approx \langle \hat{X}^{\dag}\rangle\langle \hat{X}\rangle$, thus they are prevailingly described by correlations of each dipole with all others (the corresponding terms are $N(N-1)$). The field, in turn, scales with the synchronization order parameter $X$, and thus the intracavity photon number scales with $N^2$. In this sense the regime studied in Refs. \cite{Salzburger:2004,Salzburger:2005,Salzburger:2006} is complementary to the regime analysed in this paper.\\

Finally, both models, the one of Ref. \cite{Salzburger:2006} and our model, predict stationary momentum distributions whose width is not determined by the width of the resonator. In our model, in particular, the lower bound is determined by the collective linewidth  $N\Gamma_C$ and is ultimately bound by the recoil energy in order to keep the treatment consistent with the semiclassical approximation.

\section{Conclusions}
\label{Sec:3}

In this manuscript we have analysed the semiclassical dynamics of the atomic external degrees of freedom in the parameter regime where the dipoles synchronize. We have shown that the large friction forces predicted in Ref. \cite{Xu:2016} are accompanied by the onset of an antiferromagnetic-like order, where internal and external degrees of freedom become correlated. 

Minimal temperatures are found when the parameters are chosen so that the pump rate is in the synchronization regime, $w=N\Gamma_C/4$. In this regime the incoherent pump rate indeed determines the asymptotic width of the momentum distribution. Our results suggest the possibility that sub-recoil temperature could be achieved by reducing $N\Gamma_C$, as long as this value is larger than the rate of spontaneous decay. Testing this conjecture requires a full quantum mechanical treatment of the dynamics in order to explore the ultimate limits. This is not straight-forward due to the many-body character of the laser-cooling system presented here.

\acknowledgements
The authors acknowledge discussions with Helmut Ritsch and John Cooper. This work was supported by the German Research Foundation (DACH "Quantum crystals of matter and light" and the Priority Programme 1929), by the German Ministry of Research and Education (BMBF, "Qu.com"), and by the DARPA ATN program through grant number W911NF-16-1-0576 through ARO. The views and conclusions contained in this document are
those of the authors and should not be interpreted as representing the official policies, either expressed or implied, of the U.S. Government. The U.S. Government is authorized to reproduce and distribute reprints for
Government purposes notwithstanding any copyright notation herein.

\appendix

\section{Elimination of the cavity field}
\label{App:A}

The formal solution of the Heisenberg-Langevin equation \eqref{quantuma} reads
\begin{widetext}
	\begin{align}
	\hat{a}(t)=&e^{\left(-i\Delta-\frac{\kappa}{2}\right)t}\hat{a}(0)-i\frac{g}{2}\int_{0}^{t}d\tau e^{-\left(i\Delta+\frac{\kappa}{2}\right)(t-\tau)}\sum_{j=1}^N\cos(k\hat{x}_j(\tau))\hat{\sigma}_j(\tau)+ \sqrt{\kappa}\int_{0}^{t} d\tau e^{-\left(i\Delta+\frac{\kappa}{2}\right)(t-\tau)}\hat{a}_{\text{in}}(\tau)\label{generala}\,,
	\end{align}
\end{widetext}
with $\hat{a}(0)$ the operator at time $t=0$. To eliminate the cavity field we have to substitute the operator $\hat{a}(t)$ and $\hat{a}^{\dag}(t)$ in the Heisenberg-Langevin equations for the atomic degrees of freedom (Eqs. \eqref{quantumforce}, \eqref{quantumspin}, \eqref{quantumspinz}) with an averaged field \begin{align*}
\bar{\hat{a}}(t)=\frac{1}{\Delta t}\int_{t}^{t+\Delta t} d\tau \hat{a}(\tau),
\end{align*} 
where we average over the time interval $\Delta t$.
If we now assume that $T_C\ll\Delta t\ll T_e$ and use the general expression in Eq. \eqref{generala} we derive
\begin{align}
\bar{\hat{a}}(t)\approx\frac{-i\frac{Ng}{2}\hat{X}}{\frac{\kappa}{2}+i\Delta}-\frac{-i\frac{Ng}{2}\frac{d}{dt}\hat{X}}{\left(\frac{\kappa}{2}+i\Delta\right)^2}+\hat{\mathcal{F}}(t)\,, \label{eliminateda}
\end{align}
with $\hat{X}$ the synchronization order parameter, given in Eq. \eqref{X}. We have introduced an effective Gaussian noise defined on the time scale of the atomic dynamics,
which reads 
\begin{equation}
\hat{\mathcal{F}}(t)=\sqrt{\kappa/\left(\left(\frac{\kappa}{2}\right)^2+\Delta^2\right)}\bar{\hat{a}}_{\text{in}}(t)\,,
\end{equation} 
with 
$$\bar{\hat{a}}_{\text{in}}(t)=\frac{1}{\Delta t}\int_{t}^{t+\Delta t} d\tau \hat{a}_{\text{in}}(\tau)\,.$$

The time derivative of $\hat X$ in Eq. \eqref{eliminateda} contains the non-adiabatic corrections to the cavity field, as is visible when explicitly evaluating its form:
\begin{align*}
\frac{d}{dt}\hat{X}=&\frac{1}{N}\sum_{j=1}^N\cos(k\hat{x}_j)\frac{d}{dt}\hat{\sigma}_j+\frac{1}{N}\sum_{j=1}^N\frac{d}{dt}\left(\cos(k\hat{x}_j)\right)\hat{\sigma}_j\,.
\end{align*}

\section{Numerical simulations of the semiclassical equations}
\label{App:B}

In order to perform the integration of Eqs. \eqref{sigma2}, \eqref{sigmaz2} we perform a second-order cumulant expansion and simulate the matrix $(\langle \hat{\sigma}_j^{\dag}\hat{\sigma}_l\rangle)_{1\leq j,l\leq N}$ with
\begin{align}
\frac{d}{dt}\langle\hat{\sigma}_j^{\dag}\hat{\sigma}_j\rangle=
&w(1-\langle\hat{\sigma}_j^{\dag}\hat{\sigma}_j\rangle)\nonumber\\
&+N\Gamma_C\mathrm{Im}\left(\alpha\left\langle \hat{X}^{\dag}\hat{\sigma}_j\right\rangle\right)\cos(kx_j)\,,\label{ieqj}
\end{align}
and for $l\neq j$
\begin{align}
\frac{d}{dt}\langle\hat{\sigma}_j^{\dag}\hat{\sigma}_l\rangle
\approx&
-\bigg\{w+\Gamma_C(i\alpha)\cos^2(kx_j)
\langle\hat{\sigma}_j^{\dag}\hat{\sigma}_j\rangle\nonumber\\
&+\Gamma_C(i\alpha)^*\cos^2(kx_l)
\langle\hat{\sigma}_l^{\dag}\hat{\sigma}_l\rangle\bigg\}
\langle\hat{\sigma}_j^{\dag}\hat{\sigma}_l\rangle\nonumber\\
&+N\frac{\Gamma_C}{2}(i\alpha)\cos(kx_j)
(2\langle\hat{\sigma}_j^{\dag}\hat{\sigma}_j\rangle-1)
\langle\hat{X}^{\dag}\hat{\sigma}_l\rangle
\nonumber\\
&+N\frac{\Gamma_C}{2}(i\alpha)^*\cos(kx_l)
(2\langle\hat{\sigma}_l^{\dag}\hat{\sigma}_l\rangle-1)
\langle\hat{\sigma}_j^{\dag}\hat{X}\rangle\,.\label{ineqj}
\end{align}
The simulations are performed with $N=100$ particles, the data correspond to the average taken over $1000$ trajectories. The initial state is a thermal distribution at a fixed temperature $T$ that is spatially homogeneous, with all atoms are prepared  in the excited state:
\begin{align}
\hat{\rho}_{0}=C\Pi_{j=1}^N \exp\left(-\beta\frac{p_j^2}{2m}\right)\otimes_{j=1}^N |e\rangle_j\langle e| \label{rho0}\,.
\end{align} 
Here $C$ is a normalization constant and $\beta=(k_{\text{B}}T)^{-1}$ with the Boltzmann constant denoted by $k_{\text{B}}$.

\section{Adiabatic elimination of the internal degrees of freedom in the mean-field model}
\label{App:C}

\begin{figure}[h]
	\flushleft(a)\vspace{-4ex}\\
	\center \includegraphics[width=0.75\linewidth]{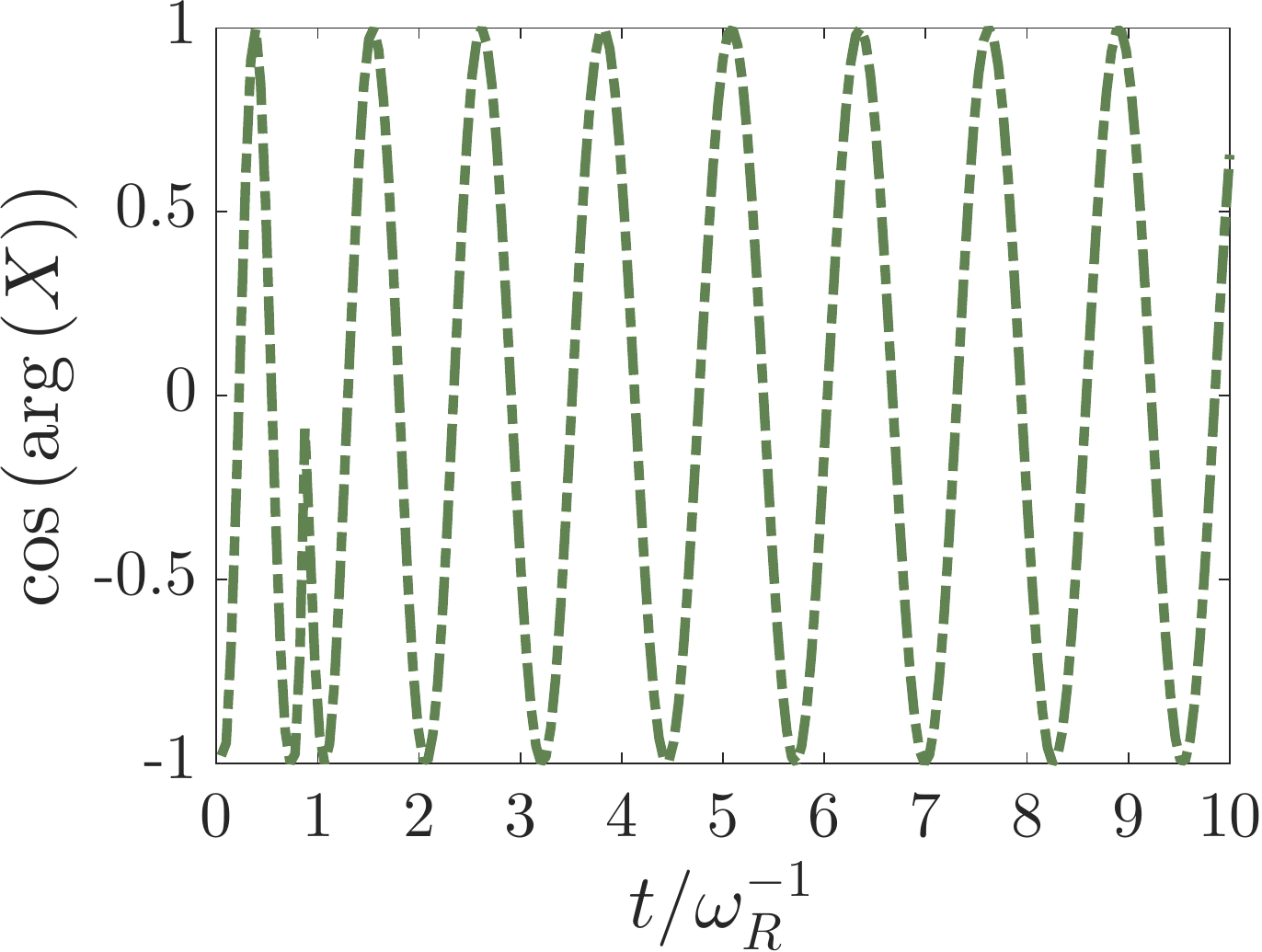}\\
	\flushleft(b)\vspace{-4ex}\\
	\center \includegraphics[width=0.75\linewidth]{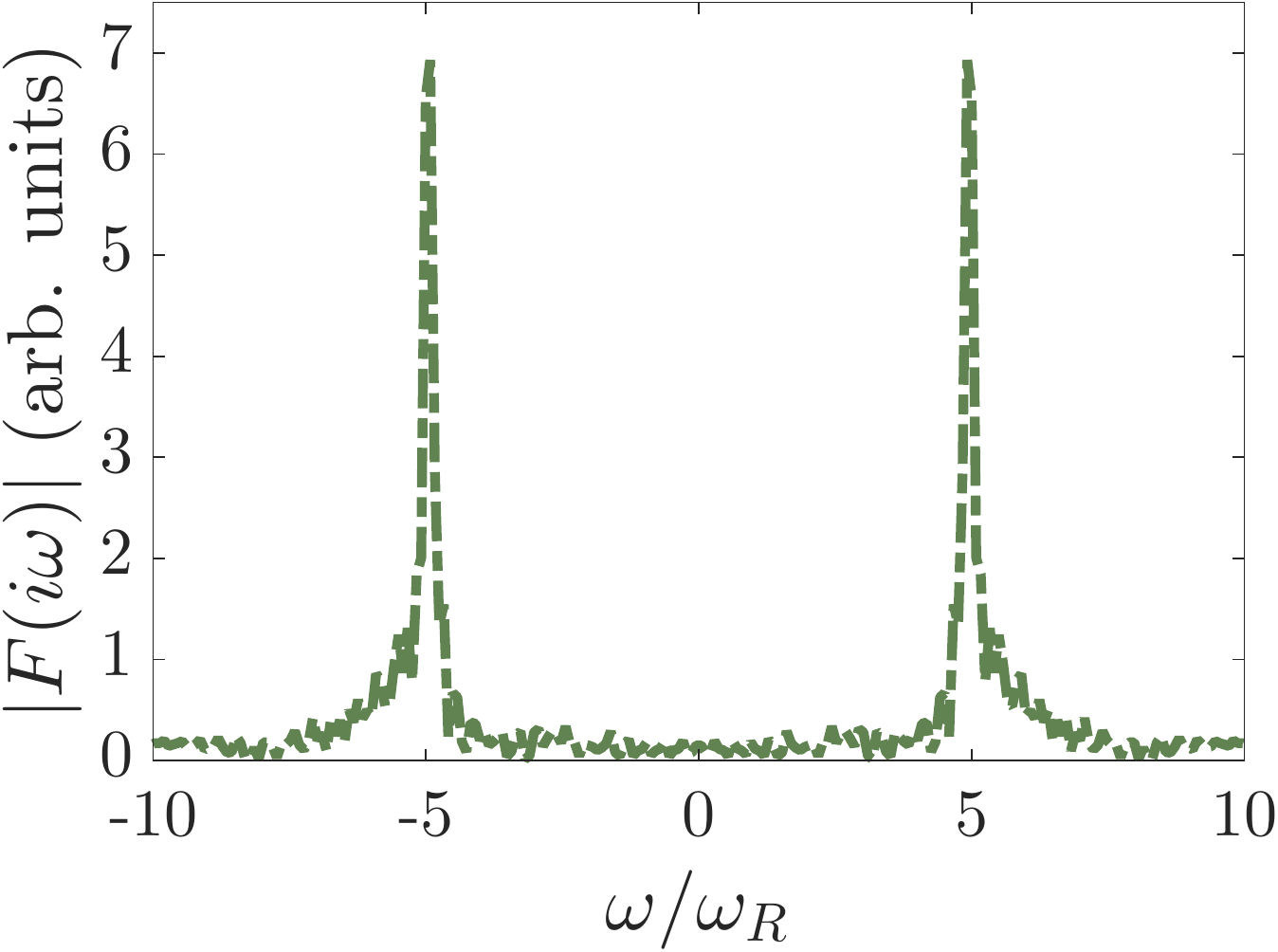}
	\caption{Dynamics of $\cos(\mathrm{arg}(X))$ as a function of time (in units of $\omega_R^{-1}$).The time evolution of the order parameter $X$ has been determined using the mean-field model for the same parameters as in Fig. \ref{N1000mean}(b). Subplot (b) shows the Laplace transform $F(i\omega)=\int_{0}^{\infty}dte^{i\omega t}\cos(\mathrm{arg}(X(t)))$ as a function of the frequency (in units of $\omega_R$). Two well defined sidebands are visible at frequency $\omega\approx\pm 5\omega_R=\pm w/2$.}
	\label{Xoscillation}
\end{figure}

In order to eliminate the internal degrees of freedom we observe that Eqs. \eqref{sj},\eqref{zj} are invariant under the transformation $s_j\to \tilde{s}_j=s_je^{-i\omega_0t}$. Performing this transformation Eqs. \eqref{sj},\eqref{zj} take the form
\begin{align}
\frac{ds_j}{dt}=&\left(i\omega_0-\frac{w}{2}\right)s_j-i\frac{N\Gamma_C}{2}\alpha^* X\cos(kx_j)z_j,\label{sj2}\\
\frac{dz_j}{dt}=&w(1-z_j)+2N\Gamma_C\mathrm{Im}\left\{\alpha X^{*}s_j\right\}\cos(kx_j)\label{zj2}.
\end{align}
We determine the stationary state by finding $\omega_0$ self consistently.
For the stationary values $s_j^{(0)}$ and $z_j^{(0)}$ of Eqs. \eqref{sj2}, \eqref{zj2}, we find
\begin{align*}
s_j^{(0)}=\frac{N\Gamma_{C}\alpha^*}{2\omega_0+iw}X\cos(kx_j)z_j^{(0)}.
\end{align*}
We apply this result to determine the synchronization order parameter by using the expression 
\begin{align*}
X=\sum_{j=1}^Ns_j^{(0)}\cos(kx_j)=\frac{N\Gamma_C\alpha^*}{2\omega_0+iw}X\sum_{j=1}^N\cos^2(kx_j)z_j^{(0)}\,.
\end{align*}
The latter can be recast in the form
\begin{align}
\frac{N\Gamma_C\alpha^*}{2\omega_0+iw}\sum_{j=1}^N\cos^2(kx_j)z_j^{(0)}=1\label{Xmeanalt}\,.
\end{align}
This expression leads to the condition $N\Gamma_{C}\alpha^*/(2\omega_0+iw)\in\mathbb{R}$, which is valid providing that
\begin{align}
\omega_0=\frac{w\Delta}{\kappa}\,. \label{omega0}
\end{align}
We numerically checked this result by calculating $\cos(\mathrm{arg}(X))$ using mean-field simulations, see Fig. \ref{Xoscillation} (a). It was observed that $X$ oscillates with a well defined frequency, as visible in
the Laplace transform of the signal, see Fig. \ref{Xoscillation} (b). For the considered parameters $w/2\approx 5\omega_R$.
With this result for $\omega_0$ it is possible to obtain the stationary states $s_j^{(0)}$ and $z_j^{(0)}$ in Eqs. \eqref{eliminatedsj} and \eqref{eliminatedzj}, respectively.

\section{Calculation of the friction force due to the coupling with the spins}
\label{App:D}
To calculate retardation effects in the elimination of the spins we replace $d/dt\to\partial/\partial t+p/m\partial/\partial x$ and identify the stationary state with
\begin{align*}
s=&s^{(0)}+\frac{p}{m}s^{(1)}\,,\\
z=&z^{(0)}+\frac{p}{m}z^{(1)}\,.
\end{align*}
We do not use the index $j$, instead we employ $x$ and $p$ since the whole approach is valid for all particles when we work in the limit $N\to\infty$. If we now use the equations for $s^{(0)}$ and $z^{(0)}$ in Eqs. \eqref{eliminatedsj} and \eqref{eliminatedzj}, this leads to the following equations for $s^{(1)}$ and $z^{(1)}$
\begin{align*}
\frac{\partial s^{(0)}}{\partial x}=&i\frac{w}{2}\alpha^*s^{(1)}-i\frac{N\Gamma_{C}}{2}\alpha^*X\cos(kx)z^{(1)}\,,\\
\frac{\partial z^{(0)}}{\partial x}=&-wz^{(1)}+2\mathrm{Im}\left(N\Gamma_{C}\alpha X^*s^{(1)}\right)\cos(kx)\,.
\end{align*}
The solutions are
	\begin{align*}
	s^{(1)}=&\xi(x)z^{(1)}\,,\\
	&+2\frac{k}{w}\frac{1}{i\alpha^*}\tan(kx)\xi(x)\frac{2|\xi(x)|^2-1}{\left(1+2|\xi(x)|^2\right)^2},\\
	z^{(1)}=&-\frac{k}{w}\tan(kx)\frac{4|\xi(x)|^2}{\left(1+2|\xi(x)|^2\right)^3}\\
	&+4\frac{k}{w}\frac{1-\left(\frac{\Delta }{\kappa/2}\right)^2}{1+\left(\frac{\Delta}{\kappa/2}\right)^2}\tan(kx)|\xi(x)|^2\frac{2|\xi(x)|^2-1}{\left(1+2|\xi(x)|^2\right)^3}.
	\end{align*}

\section{Calculation of the diffusion coefficient}
\label{App:E}

To calculate the diffusion coefficient we use \cite{Nienhuis}
\begin{align}
2D=\int_{0}^{\infty}d\tau\bigg(&\frac{1}{2}\left\langle F(0)F(\tau)+F(\tau)F(0)\right\rangle_{\hat{\rho}_{st}}\nonumber\\&-\left\langle F(0)\right\rangle_{\hat{\rho}_{st}}\left\langle F(\tau)\right\rangle_{\hat{\rho}_{st}}\bigg)\,,\label{regression}
\end{align}
where the force $$F(\tau)=\frac{\hbar k}{2}\sin(kx)w\xi^*\hat{\sigma}(\tau)+\mathrm{H.c.}$$ is obtained after adiabatic elimination of the cavity (see Eq. \eqref{classicalforce}). The expectation values are calculated with the stationary density matrix $\hat{\rho}_{\mathrm{st}}$ and $\left\langle \,.\,\right\rangle_{\hat{\rho}_{st}}=\mathrm{Tr}_i\left(\,.\,\hat{\rho}_{\mathrm{st}}\right)$ where $\mathrm{Tr}_i$ is the trace over the internal degrees of freedom. As we can see in Eqs. \eqref{sigma2}, \eqref{sigmaz2} the motion of $\hat{\sigma}$, $\hat{\sigma}^{\dag}$, and $\hat{\sigma}^z$ is coupled. We define the vector
\begin{align*}
v=&\begin{pmatrix}
\hat{\sigma}\\
\hat{\sigma}^{\dag}\\
\hat{\sigma}^{z}
\end{pmatrix}\,,\\
\end{align*}
and write the equations of motion for the spins as
\begin{align}
\frac{dv}{dt}=&\Omega v+b+\mathcal{S}\label{vdynamic}\,,
\end{align} 
where the matrix $\Omega$ is defined as
\begin{align*}
\Omega=&\begin{pmatrix}
i\frac{w}{2}\alpha^*&0&-i\frac{w}{2}\alpha^*\xi\\
0&-i\frac{w}{2}\alpha&i\frac{w}{2}\alpha\xi^*\\
-iw\alpha\xi^*&iw\alpha^*\xi&-w
\end{pmatrix}\,,\\
\end{align*}
and the vector $b$ reads
\begin{align*}
b=&\begin{pmatrix}
0\\0\\w
\end{pmatrix}\,.
\end{align*}
The noise $\mathcal{S}(\tau)=(S^{-}(\tau)\,\, S^{+}(\tau)\,\,S^{z}(\tau))^T$ is defined by
\begin{align*}
\left\langle \mathcal{S}(\tau)\mathcal{S}^T(\tau')\right\rangle=&\begin{pmatrix}
0&0&0\\
1&0&-2\langle\hat{\sigma}^{\dag}(\tau')\rangle_{\hat{\rho}_{st}}\\
-2\langle\hat{\sigma}(\tau')\rangle_{\hat{\rho}_{st}} &0&2(1-\langle\hat{\sigma}^{z}(\tau')\rangle_{\hat{\rho}_{st}})
\end{pmatrix}\delta(\tau-\tau')\,.
\end{align*}
The formal solution of Eq. \eqref{vdynamic} is
\begin{align}
v(\tau)=e^{\Omega \tau}v+(e^{\Omega\tau}-1)\Omega^{-1}b+\int_{0}^{\tau}d\tau'e^{\Omega (\tau-\tau')}\mathcal{S}(\tau')\label{solutionv}\,.
\end{align}
If we apply the limit $\tau\to\infty$ for Eq. \eqref{solutionv} we get the stationary states for $\hat{\sigma}$, $\hat{\sigma}^{\dag}$, and $\hat{\sigma}^{z}$ which define the stationary density matrix 
\begin{align}
\hat{\rho}_{\mathrm{st}}=\begin{pmatrix}
\frac{1+|\xi|^2}{1+2|\xi|^2}&\frac{\xi^*}{1+2|\xi|^2}\\
\frac{\xi}{1+2|\xi|^2}&\frac{|\xi|^2}{1+2|\xi|^2}
\end{pmatrix}\,. \label{stationarydensity}
\end{align}
If we now use the general solution (Eq. \eqref{solutionv}) together with the density operator in Eq. \eqref{stationarydensity} to calculate the diffusion coefficient (Eq. \eqref{regression}) we obtain 
\begin{widetext}
	\begin{align}
	2D=&\frac{(\hbar k)^2}{2}w\tan^2(kx)|\xi|^2\left(1+\frac{2\left(\frac{\Delta_c}{\kappa/2}\right)^2|\xi|^2}{1+2|\xi|^2}-2\frac{\left(\frac{\Delta_c}{\kappa/2}\right)^2}{1+\left(\frac{\Delta_c}{\kappa/2}\right)^2}\frac{|\xi|^2}{\left(1+2|\xi|^2\right)^2}\frac{5+\left(\frac{\Delta_c}{\kappa/2}\right)^2+4\left(\left(\frac{\Delta_c}{\kappa/2}\right)^2+1\right)|\xi|^2}{1+2|\xi|^2}\right)\,. \label{Diffusioncoefficient}
	\end{align}
	\end{widetext}

\section{Detailed comparison with the results of Ref. \cite{Salzburger:2006}}
\label{App:F}

An interesting example highlighting the complementarity of the two approaches is found by comparing the expectation value of population inversion. In the adiabatic limit (in the thermodynamic limit $Ng^2=$const) the population inversion we calculate reads
\begin{align*}
\langle z\rangle\approx\left\langle\frac{1}{1+2|\xi(x)|^2}\right\rangle\,,
\end{align*}
which implies that there is always population inversion $\langle z\rangle\geq0$. Note that if all particles are in the excited state then $\langle z\rangle=1$, thus $X=0$ and the atoms are not synchronized. Hence synchronization requires a non-vanishing expectation value of the dipole. This is not the case for the parameters of Ref. \cite{Salzburger:2006}, where the expectation value of the dipole vanishes at steady state. 

We now show that also in Ref. \cite{Salzburger:2006} all particles are in the excited state if one takes the equations of motion there defined, neglects spontaneous emission and performs the limit $N\to\infty$ with $Ng^2=$ constant, taking $\kappa$ as the largest parameter.  For this purpose we take the formula for the population inversion $z_N$ in Eq. (25) of Ref. \cite{Salzburger:2006} , and report it using our notations giving
\begin{align}
z_N=&\frac{\kappa w+N\Gamma w-w\sqrt{\left[\kappa +N\Gamma  \right]^2-4\kappa N \Gamma}}{2N\Gamma w}\,. \label{zN}
\end{align} 
We already neglected spontaneous emission $\gamma=0$ and applied the limit $g^2\propto N^{-1}$ with $N\to\infty$. In Eq. \eqref{zN} the frequency $\Gamma$ is the emission rate  defined as $\Gamma=wg^2/(w^2+\Delta^2)$. From this expression one can verify that when $\kappa$ is chosen to be the largest frequency then $z_N=1$.


\begin{thebibliography}{99}	

\bibitem{Wineland:1979}
D. J. Wineland and W. M. Itano,
Phys. Rev. A {\bf 20}, 1521 (1979). 


\bibitem{Chu:1998}
S. Chu, Rev. Mod. Phys. {\bf 70}, 685 (1998); C. N. Cohen-Tannoudji,
Rev. Mod. Phys. {\bf 70}, 707 (1998); W. D. Phillips,
Rev. Mod. Phys. {\bf 70}, 721 (1998).

\bibitem{Wineland:1999}
C. E. Wieman, D. E. Pritchard, and D. J. Wineland,
Rev. Mod. Phys. {\bf 71}, S253 (1999).

\bibitem{Metcalf:2003}
H. J. Metcalf and P. van der Straten, {\it Laser Cooling and Trapping} (Springer, New York, 1999).

\bibitem{Aspelmeyer}
M. Aspelmeyer, T. J. Kippenberg, and F. Marquardt, Rev. Mod.
Phys. {\bf 86}, 1391 (2014).

\bibitem{Rubinztein}
A. Rayner, N. R. Heckenberg, and H. Rubinsztein-Dunlop,
J. Opt. Soc. of Am. B {\bf 20}, 1037 (2003).

\bibitem{Eschner:2003}
J. Eschner, G. Morigi, F. Schmidt-Kaler, and R. Blatt, J. Opt. Soc. Am. B {\bf 20}, 1003
(2003).

\bibitem{Walker:1990}
T. Walker, D. Sesko, and C. Wieman, Phys. Rev. Lett. {\bf 64}, 408 (1990).
	
\bibitem{Marksteiner:1996}
S. Marksteiner, K. Ellinger, and P. Zoller, Phys. Rev. A {\bf 53}, 3409 (1996).

\bibitem{Castin:1998}
Y. Castin, J. I. Cirac, and M. Lewenstein, Phys. Rev. Lett. {\bf 80}, 5305 (1998).

\bibitem{Cirac:1996}
J. I. Cirac,  M. Lewenstein,  and  P. Zoller,  Europhys.  Lett. {\bf 35}, 647 (1996).

\bibitem{Grimm}
S. Stellmer, B. Pasquiou, R. Grimm, and F. Schreck,
Phys. Rev. Lett. {\bf 110}, 263003 (2013).

\bibitem{Horak:1997}
P. Horak, G. Hechenblaikner, K. M. Gheri, H. Stecher, and H. Ritsch, Phys. Rev. Lett. {\bf 79}, 4974 (1997).
	
\bibitem{Vuletic:2000}
V. Vuletic and S. Chu, Phys. Rev. Lett. {\bf 84}, 3787 (2000).
	
\bibitem{Ritsch:RMP}
H. Ritsch, P. Domokos, F. Brennecke, and T. Esslinger, Rev. Mod. Phys. {\bf 85}, 553 (2013).

\bibitem{Vuletic:preprint}
M. Hosseini, Y. Duan, K. M. Beck, Y.-T. Chen, and V. Vuletic, preprint arXiv:1701.01226.
	
\bibitem{Domokos:2002}
P. Domokos and H. Ritsch, Phys. Rev. Lett. {\bf 89}, 253003 (2002).
	
\bibitem{Asboth:2005}
J. K. Asb{\'o}th, P. Domokos, H. Ritsch, and A. Vukics, Phys. Rev. A {\bf 72}, 053417 (2005).
	
\bibitem{Schuetz:2014}
S. Sch\"utz and G. Morigi, Phys. Rev. Lett. {\bf 113}, 203002 (2014).	

\bibitem{Schuetz:2015}
S. Sch\"utz, S. B. J\"ager, and G. Morigi,  Phys. Rev. A {\bf 92}, 063808 (2015).
	
\bibitem{Salzburger:2004}
T. Salzburger and H. Ritsch, Phys. Rev. Lett. {\bf 93}, 063002 (2004).
		
\bibitem{Salzburger:2005}
T. Salzburger, P. Domokos, and H. Ritsch, Phys. Rev. A {\bf 72}, 033805 (2005).
		
\bibitem{Salzburger:2006}
T. Salzburger and H. Ritsch, Phys. Rev. A {\bf 74}, 033806 (2006).

\bibitem{Bohnet:2012}
J. G. Bohnet, Z. Chen, J. M. Weiner, D. Meiser, M. J. Holland, and J. K. Thompson,
Nature (London) {\bf 484} ,78 (2012).

\bibitem{Xu:2016}
M. Xu, S. B. J\"ager, S. Sch\"utz, J. Cooper, G. Morigi, and M. J. Holland, Phys. Rev. Lett. {\bf 116}, 153002 (2016).

\bibitem{Meiser:2009}
D. Meiser, J. Ye, D. R. Carlson, and M. J. Holland
Phys. Rev. Lett. {\bf102}, 163601 (2009).
	
	\bibitem{Meiser:2010:1}
	D. Meiser and M. J. Holland, 
	Phys. Rev. A {\bf81}, 033847 (2010).

\bibitem{Gardiner:1985}
C. W. Gardiner and M. J. Collett, Phys. Rev. A {\bf 31}, 3761 (1985).


\bibitem{Stenholm:1986}
S. Stenholm, Rev. Mod. Phys. {\bf 58}, 699 (1986).

\bibitem{SDE}
For details on the implementation of the stochastic differential equations in a similar setup see S. Sch\"utz, H. Habibian, and G. Morigi, Phys. Rev. A {\bf 88}, 033427 (2013).		


\bibitem{CohenTannoudji:Book}
C. Cohen-Tannoudij, J. Dupont-Roc, G. Grynberg, {\it Atom-
Photon Interactions} (Wiley, Toronto, 1992).
		
		
\bibitem{Nienhuis}
G. Nienhuis, P. van der Straten, and S-Q. Shang, Phys. Rev. A {\bf44}, 462 (1991).
	
	
\end{thebibliography}
\end{document}